\documentclass{jfm}
\usepackage{graphicx}
\usepackage[section]{placeins}
\usepackage{epstopdf, epsfig}
\usepackage{enumitem}

\usepackage{amssymb}
\usepackage{amsmath}
\usepackage{soul}
\usepackage{xcolor}
\usepackage{textcomp}

\shorttitle{Drop impact on a continuous liquid jet}
\shortauthor{Baumgartner, Bernard, Weigand, Lamanna, Brenn and Planchette}

\title{{Influence of liquid miscibility and wettability 
on the structures produced by drop-jet collisions}}

\author{David Baumgartner\aff{1}, Ronan Bernard\aff{2}, Bernhard Weigand\aff{2}, Grazia Lamanna\aff{2}, 
  G\"unter Brenn\aff{1}
 \and Carole Planchette\aff{1} \corresp{\email{carole.planchette@tugraz.at}}}

\affiliation{\aff{1}Institute of Fluid Mechanics and Heat Transfer, Graz University of Technology, 8010 Graz, Austria
\aff{2}Institute of Aerospace Thermodynamics, University of Stuttgart, 70569 Stuttgart, Germany}

\begin{document}

\maketitle

\begin{abstract}

Collisions between a stream of drops and a continuous jet of a different liquid are experimentally investigated. In contrast to previous studies, our work focuses on the effects of liquid miscibility and wettability on the collision outcomes. Thus,  miscible and immiscible liquids providing total and partial wetting are used. We show that, as long as  the jet surface tension is smaller than the drop surface tension, the drops can be encapsulated by the jet, providing the so-called drops-in-jet structure. The transitions between the different regimes remain similar in nature with a capillary fragmentation responsible for the jet break-up and an inertial fragmentation causing the drops (and then possibly the jet) to break up. The dimensionless numbers  proposed in the literature to model the inertial fragmentation thresholds do not bring the results obtained with different liquids at the same critical value. We explain the reason via a detailed analysis of the collisions, accounting for the drop and jet extensions and their kinetics. The drop fragmentation is found to occur during the recoil phase, leading us to propose a new dimensionless parameter that successfully reproduced all our experimental data obtained with immiscible liquids. Finally we demonstrate that the most dramatic change of the collision outcomes is produced by using drops that totally wet the jet. In this case, the encapsulation of the drops cannot be achieved, constituting a true limit to some applications based on the solidification of the drops-in-jet structure. 
\end{abstract}


\section{Introduction}\label{intro}

Drop impacts and collisions are widely encountered in natural phenomena like rain and are commonly used in a broad range of industrial processes such as  spray coating, printing, encapsulation and formulation of food, cosmetic or pharmaceutical products. While originally mainly drop  collisions involving a single liquid have been studied, several subsequent studies have been focusing on drop collisions where more than one liquid is used \citep{Rein1993, Berberovic2009, Gao2005, Yarin2006}. These studies may have been motivated by basic research  but are also often of strong practical interest. Drop impacts onto liquid pools may  be important for emulsification and de-emulsification and therefore related processes as various as mixing and de-pollution \citep{Kavehpour_2014} or for encapsulation technologies \citep{BRANDENBERGER199873, Serp2000, Haeberle2008}. Drop impact onto liquid films has gained considerable interest due to its relevance to lubrication and combustion, especially considering diesel drops impacting on a thin film of oil as it happens at the piston wall of a thermal engine \citep{Josserand2003, Wang2000, Cossali1997, Geppert_2017}. Such impacts are also of importance to the printing and coating industries where  droplets may impact on  deposited not yet cured ink layers \citep{Martin_2008}. In the last decades, drop-drop collisions using more than one liquid have been considered. Many different applications were named such as the use of the combined drop as a micro-reactor \citep{Teh2008}, the possibility to improve combustion \citep{KADOTA_2002, Wang_2004, Tsuru_2019}, the possibility to manufacture capsules \citep{Yeo_Chen_Basaran_Park_2004} or more generally the use of such collisions as a method to encapsulate  one liquid in the shell of another one \citep{Planchette2012}. Recently, interest has also raised for drop-jet collisions \citep{Planchette2018, PlanchetteBrenn2018} which has also been called \textit{in-air microfuidics} \citep{Visser2018}. Such collisions typically  aim - beside gaining basic knowledge on capillaro-inertial systems - at producing advanced capsules or fibers by solidifying the liquid structures formed upon collisions \citep{Kamperman2018}.

Yet, most of these studies focus on the limit between different regimes such as deposition/splash; encapsulation/fragmentation; coalescence/resting \citep{Yarin2006, Kavehpour_2014, Lhuissier2013} %leaving the questions about the spatial liquid distribution  mainly untouched. 
without specifying where (in which fragment) and under which form (single or multiple droplets, totally or partially encapsulated) each liquid can be found.  Indeed, considerable advances have been made in understanding the limit of splashing and deposition for drop impacts onto liquid films of various thicknesses. Especially the role of the viscosity ratio which has been intensively studied both experimentally and numerically \citep{Kittel_Roisman2018, Chen_2017, Berberovic2009, josserand2016, Schroll_Zaleski2010} became clearer. Similarly, for drop-drop collisions, both experimental and numerical studies have drawn the main trends and demonstrated that the effects of the liquid viscosity are significantly different in the encapsulated and encapsulating phases \citep{Roisman2004, Planchette2012, PLANCHETTE201089,Weigand2007}. Yet, the effects of liquid wettability and miscibility are not  addressed,  except when liquids are gently brought into contact \citep{SANTIAGOROSANNE2001375,Blanchette_2009},  which irremediably leads after some time to the thermodynamic equilibrium. %Further, the expected spatial distribution of the liquids after the collision and its evolution due to changes in wettability and miscibility are barely studied. We believe  this point to be very critical. 
We believe that the expected spatial distribution of the liquids after  collision, and the effects of wettability and miscibility on it, are very critical. 
Indeed, considering drop impacts onto a liquid pool, drop impacts onto liquid films, drop-drop collisions or drop-jet collisions, the question remains: where is the drop liquid after impact? When the liquids are immiscible, %is it at one place only in the form of one unique entity or not? W
what can be expected from reversing the two liquids? Do miscible liquids give qualitatively the same kinds of regimes and are the transitions modified? These questions arise for any application and are of basic interest for any capillary-inertial (possibly viscous) system.

In this paper, we make a first step toward a complete understanding and modelling of such systems by studying the effects of liquid miscibility and wettability on the outcomes of drop-jet collisions. Note that, drop impacts onto liquid films using identical liquid pairs have also been carried out. For concision, the drop-film impact results will be reported in an independent article and put in relation with those obtained for drop-jet collisions and presented here. In both cases, the study is experimental. Beside  recent considerable progress of numerical simulation \citep{Dai2005,Mazloomi2016,liu_zhang_gao_lu_ding_2018,LUNKAD20077214}  huge difficulties and associated large computational costs remain. The challenge lies in numerically accounting  for three-phase flows (interface tracking, reconstruction) with moving contact lines and the adequate spatial and time resolutions \citep{bazhlekov_shopov_1997,Sui2014}. Indeed, the typical spatial scale ranges from the thickness of very thin liquid films (spreading film, lamella, air cushion separating the drops) to the diameter of the whole drops. The temporal evolution is very fast due to the small dimensions ($100$ \textmu m) and the large relative velocity ($10$ m/s),  but the system must be followed over several ms to identify the collision outcomes. The viscosity and density differences between the surrounding gaseous phase and the two liquid phases adds complexity, especially in terms of simulation stability. Further, the drop-jet system involves the repetition of drop impacts with a typical frequency of $10$ kHz and the presence of a continuous jet, thus constituting an open system. For all these reasons, the experimental approach is for such complex collisions the most efficient one.
To our knowledge, the reported studies of drop-jet collisions, all experimental,  remain very rare. Apart from the seminal work by \cite{Chen2006_dropjet}, which uses the same liquid for the jet and the drops, other studies have employed different liquids. In \cite{Planchette2018, PlanchetteBrenn2018}, the focus was on immiscible liquids with the jet liquid totally wetting the drop liquid. The aim of these two studies lies in describing the obtained structures (characteristic for different regimes), as well as in  predicting the transition(s) between these regimes. The liquids were varied to vary the viscosity of both phases, but the total wetting condition was not changed. Finally, recent works  \citep{Visser2018, Kamperman2018} have focused on the solidification of the obtained structures. For this purpose, miscible yet different liquids were used, but the effects of the liquid properties on the collision outcomes was not addressed. We believe this is a key-point and potential bottle-neck for further applying this encapsulation method and have therefore oriented our present study toward the effects of liquid miscibility and wettability onto the collision outcome, and especially onto the spatial distribution of the liquid after the collisions. Thus, the drop and jet diameters are fixed, % in this study to $200$ \textmu m and $300$ \textmu m, respectively.
and only head-on collisions are studied. Further, the  tangential component of the relative drop-jet velocity is kept below $10$ \% of the total relative velocity in order to limit geometric effects as well as additional shear.

Aiming to characterize and understand the influence of liquid wettability and miscibility, we have carefully  selected several liquid pairs. This selection, detailed in section \ref{liq}, comprises both miscible and immiscible liquids. The immiscible liquids probe total and partial wetting of the drops by the jet as well as total wetting of the jet by the drops. For all these combinations, we report the collision outcomes and discuss their classification in the form of regimes. The regimes are elaborated considering either the fragmentation of any of the two phases (for immiscible liquids)  or the spatial distribution of the two liquids after the collisions (for miscible liquids). % OLD to be removed % We show very strong similarities between miscible and immiscible liquids which broaden the application field of such collisions to miscible liquids. In contrast with this finding, we demonstrate that dramatic differences are caused by reversing the drop and jet liquid with immiscible totally wetting liquids. In this case, the outcomes change in nature and compromise the usage of such collisions as an encapsulation method. 
Beyond the classification of the collision outcomes, we study the transitions between the observed regimes. 
%NEW below
The inertial fragmentation being not universally predicted by the existing models \citep{Planchette2018, PlanchetteBrenn2018}, a detailed analysis of  our results is conducted. Considering both energy-based arguments and kinetic ones,  %\citep{hoepffner2013,ilass_2019, hoath_2013},   
a criterion for inertial fragmentation is established and an associated parameter derived.

The paper is organized as follows: first the experimental methods and materials are presented, followed by the results obtained with our "reference system".% made of aqueous drops and a continuous jet of silicone oil. 
These results are then compared to the other configurations starting with miscible liquids and followed by the reversed reference configuration. %Similarities and differences are underlined and corresponding interpretations are given. 
Results obtained with immiscible liquids providing both partial and total wetting are analyzed in details revealing the drop and jet extension/recoil and associated kinetics. % as well as the  drop fragmentation and its mechanism.%is then carefully investigated and is shown to happen during the recoil phase of both the encapsulated drop and the jet, following a typical end-pinching mechanism. 
Combining these findings with observations of the drop fragmentation, a new  dimensionless parameter is established whose relevance is demonstrated by testing against the experimental results.
The paper ends with conclusions.

%The outcome of drops impacting on an immiscible or miscible continuous liquid jet is influenced by many different physical effects, such as energy conversion, viscous dissipation, the influence of thermodynamics of interfaces, but also fluid properties and ambient conditions are important parameters. Since this study mainly focuses on the effect of miscibility and wettablility, the first series of experiments includes G50 as drop liquid (coloured with blue dye) and transparent  oil as jet liquid. According to surface tension and interfacial tension total wetting is the first important boundary condition. In order to cover a broad range of collision outcomes the relative velocity $U$ is adjusted between $2 m/s$ and $10 m/s$ and the collision angle $\alpha$ varies between $15^\circ$ and $60^\circ$. One can expect, by taking the positive spreading parameter $S$ into account, that silicone oil immediately spreads around G50 droplets in order to reduce its surface energy \citep{Gruyter2017}. Further, different collision outcomes may be expected by increasing the kinetic energy of the impacting drop, such as drop fragmentation, jet fragmentation or fragmentation of both.

\section{Experimental methods and materials}\label{metandmat}

\subsection{Experimental set-up}\label{method}
\begin{figure}
  \centering 
  \includegraphics[width=\textwidth]{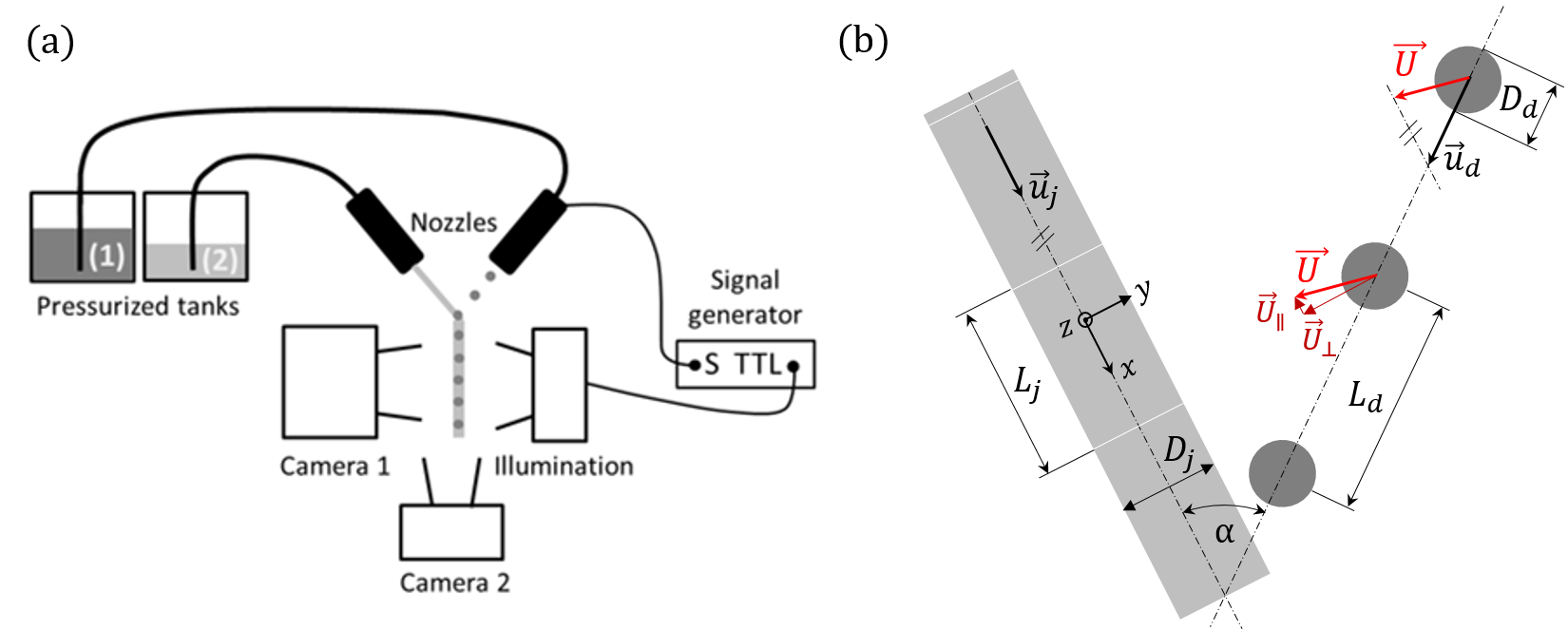}
  \caption{Adapted from \cite{PlanchetteBrenn2018}. (a) Sketch of the experimental set-up used to produce and image drop-jet collisions; (b) Sketch representing the geometric and kinetic parameters of a collision.}\label{fig:setup}
\end{figure}

The present study focuses on head-on collisions of monodisperse droplets with a continuous liquid jet. Head-on collisions are  collisions for which 
the trajectory of the droplet and the one of the jet are in the same plane. 
To accurately produce such collisions, the experimental set-up sketched in figure \ref{fig:setup}(a) is used. A droplet generator \citep{Brenn-Durst-Tropea_1996} and a nozzle  producing a stream of droplets and a continuous jet respectively, are fixed on micro traverses enabling the accurate alignment of their trajectories. Different liquids are used to produce the drops and the jet which are supplied by two different pressurized tanks. Nozzle orifices are chosen to obtain droplet diameters in the range of $D_d=200\pm{20}\,$\textmu m and jet diameters of $D_j=290\pm{20}\,$\textmu m. Here and in the rest of the article, subscript \textit{d} is used for the drop parameters or the drop liquid properties, whereas \textit{j} refers to the parameters and properties related to the jet and jet liquid.  A signal generator supplies both  the droplet generator and a stroboscopic light (LED) with the same frequency $f_d$ 
(8000 Hz $<\,f_d\,<$ 28000 Hz)\,\,enabling the record of frozen collision pictures. For a given collision, pictures are  simultaneously recorded with two  cameras (1 and 2) providing orthogonal and front views. The drops are darkly dyed producing a contrast with the clear transparent jet, and therefore providing the spatial distribution of the two liquids during and after the collisions. The public domain software ImageJ (https://imagej.nih.gov/ij/) is used to post-process the pictures and calculate the collision parameters, which are shown in figure \ref{fig:setup}(b). These parameters comprise both drop and jet diameter,  $D_d$ (180 to 220 $\mu m$)  and $D_j$ (270 to 310 $\mu m$), respectively; the spatial periods of the drop $L_d$ (300 to 1000 $\mu m$) and of the jet $L_j$ (300 to 750 $\mu m$); the collision angle $\alpha$ (15 to 60$^{\circ}$); and the velocity of drop $\vec{u}_d$ (4 to 15 $m.s^{-1}$) and jet $\vec{u}_j$ (3 to 14 $m.s^{-1}$). The spatial period $L_j$ corresponds to the distance advected between two consecutive impacts, i.e. $L_j=u_j/f_d$. The relative velocity $\vec{U}$ between the drops and the jet is simply calculated with the formula $\vec{U}=\vec{u}_d-\vec{u}_j$ and typically ranges between 2 and 9 $m.s^{-1}$. The typical uncertainties are at maximum of 5\% for the dimensions and velocities at stake. The method used to obtain the velocities is presented in appendix \ref{appD}.

Note that, in contrast to drop-drop collisions, the orientation of the relative velocity with respect to the jet trajectory also plays a role. Indeed, the relative velocity can be decomposed in two components  $U_\parallel = u_d\,$cos$(\alpha) - u_j$ in jet direction and $U_\perp = u_d\,$sin$(\alpha)$ orthogonal to the jet. In the present study the parallel component  $U_{\parallel}$ of the relative velocity $U$ is set to zero (practically $U_{\parallel}<0.1\,U $) by adjusting $u_j$, $u_d$ and $\alpha$. This adjustment enables a purely orthogonal impact on the jet and  reduces potential shearing effects caused by the parallel component of the relative velocity.\\

%\FloatBarrier

\subsection{Liquids}\label{liq}

The density $\rho$,  dynamic viscosity $\mu$, surface and interfacial tensions $\sigma$ are the relevant liquid properties for drop-jet collisions. The density is measured by weighing a liquid volume of $100\,$ml, the viscosity is determined with a glass capillary viscometer, and the surface tension is determined with the help of the pendant drop method.  Note that for interfacial tensions $\sigma_{d/j}$ the values are either measured with the pendant drop method (static values) or taken from the literature (dynamic values), \citep{Georgie2018, TAKAMURA201250}. This only concerns  n-hexadecane and is indicated by  $^{(1)}$  in table  \ref{tab:fluidproperties} listing all liquid properties.  %Yoon1992 for perfluo
%Peters2013 for silicon oil
This choice is due to the fact that n-hexadecane/water interfacial tension is known to be sensitive to polar contaminants  which play no role on short time scales (typically some minutes, to be compared to 1 ms, the maximum time scale of our observations) but influence the static measurement available via the in-house pendant drop device \citep{kralchevsky_2019}.  Yet, no significant differences are expected, and these values are mainly used to distinguish between total and partial wetting, which is further confirmed by macroscopic observation following the evolution of an oil drop gently deposited on the surface of a liquid bath made of the aqueous glycerol solution (movies available online \cite{sup_mat}).
 
 In this study, the liquids are selected to miscibility and wettability independently from other liquid properties.  Thus, for all situations,  the viscosity remains similar, so that mainly the surface and interfacial tensions are varied, except for the perfluorodecalin whose density is twice as large as for the other liquids.
 More precisely, we use a reference system made of silicone oil (Carl Roth GmbH, Germany) for the jet and an aqueous solution of glycerol ($\geq98\,$\%, Carl Roth GmbH, Germany) at 50$\,$\%w:w for the drops.  %The properties of this liquid pair are  listed in table \ref{tab:fluidproperties}.
  As predicted by the measured spreading parameter $S=\sigma_d-(\sigma_{d/j}+\sigma_j) \approx+14\,$mN/m, and in accordance with the literature \citep{Ross1992, Capillarity_2004}, the silicone oil (SO 5) totally wets the aqueous solution of glycerol, denoted G5 in the rest of the article. 
  
To study the effects of liquid miscibility, an aqueous solution of ethanol and glycerol (noted EtOH) is prepared that approaches the properties of the silicone oil. %, see table \ref{tab:fluidproperties}. 
Keeping G5 drops and substituting for the jet SO 5 by EtOH provide a pair of miscible liquids whose interfacial tension is thus zero. 
The aqueous ethanol solution  consists of ethanol ($\geq99.5\,$\%, Carl Roth GmbH, Germany), glycerol ($\geq98\,$\%, Carl Roth GmbH, Germany) and deionized water with the mass composition of 55$\,$\%, 30$\,$\% and 15$\,$\%, respectively.    %      For this liquid pair, 

For better understanding the effects of the relative liquid wettability and its potential limitation in terms of encapsulation, the silicone oil is substituted by n-hexadecane (ReagentPlus 99$\,$\%, Sigma-Aldrich, USA). Alkanes are well known to provide partial wetting with aqueous solutions, which is confirmed by the evaluation of the spreading parameter  $(S\approx-10\,$mN/m$\,<0)$ and by the observation of n-hexadecane lenses on top of G5. %Thus, this liquid pair is chosen to vary the wettability leaving the other parameters unchanged. 
Finally and in order to better evaluate and verify the scalings proposed later in this article, we use a fluorinated oil, perfluorodecalin (Apollo Scientific Ltd, UK). %As it can be seen from table \ref{tab:fluidproperties}
The main difference to our reference system is the oil density.

When performing the experiments the drop liquid is always coloured, the jet liquid  remaining uncoloured. For aqueous solutions, the dye is Indigotin 85 (E 132, BASF, Germany). Novasolve Blue 298 (Nova Specialty Chemicals LLC, USA) is used to colour silicone oil. 

\begin{table}
\begin{center}
\def~{\hphantom{0}}
  \begin{tabular}{lcccc}
      $Liquid $ & $Density$ & $Dynamic$ &  $Surface$& $Interfacial$\\
    (abbreviation) & $\rho$ & $viscosity$ &  $tension$& $tension$\\
        &  [g/dm$^3$] & $\mu$ [mPas] &  $\sigma$ [mN/m] & $\sigma_{d/j}$ [mN/m] \\[3pt]
       Glycerol 50\% (G5)   & $1116\pm{2}$ & $4.97\pm{0.1}$  & $68\pm{2}$ & -\\
       Silicone oil (SO 5)   & $908\pm{5}$ & $5.1\pm{0.05}$  & $19.5 \pm{0.5}$  & $34\pm{1}$\\
       n-hexadecane  & $767\pm{10}$ & $3.5\pm{0.3}$ & 26.5$\pm{1}$ & 50$\pm{2} ^{(1)}$\\
       Ethanol 55\% (EtOH) & $936\pm{10}$ & $4.58\pm{0.4}$ & 25.7$\pm{0.7}$ & 0\\
       Perfluorodecalin & $1908\pm{5}$ & $5.9\pm0.5$ &$13\pm{1}$  & 36$\pm{0.4}$\\
  \end{tabular}
  \caption{Properties of the liquids used in the present study. (1) indicates values taken from the literature against pure water or against decane after extrapolation to the studied system.}
  \label{tab:fluidproperties}
  \end{center}
\end{table}

All experiments, including the measurements of fluid properties, are carried out at the same ambient temperature $T_{amb}=23.5\pm{1}{^\circ}$C.

\subsection{Image and shape analysis}\label{sec:image}

\begin{figure}
  \centering 
  \includegraphics[width=\textwidth]{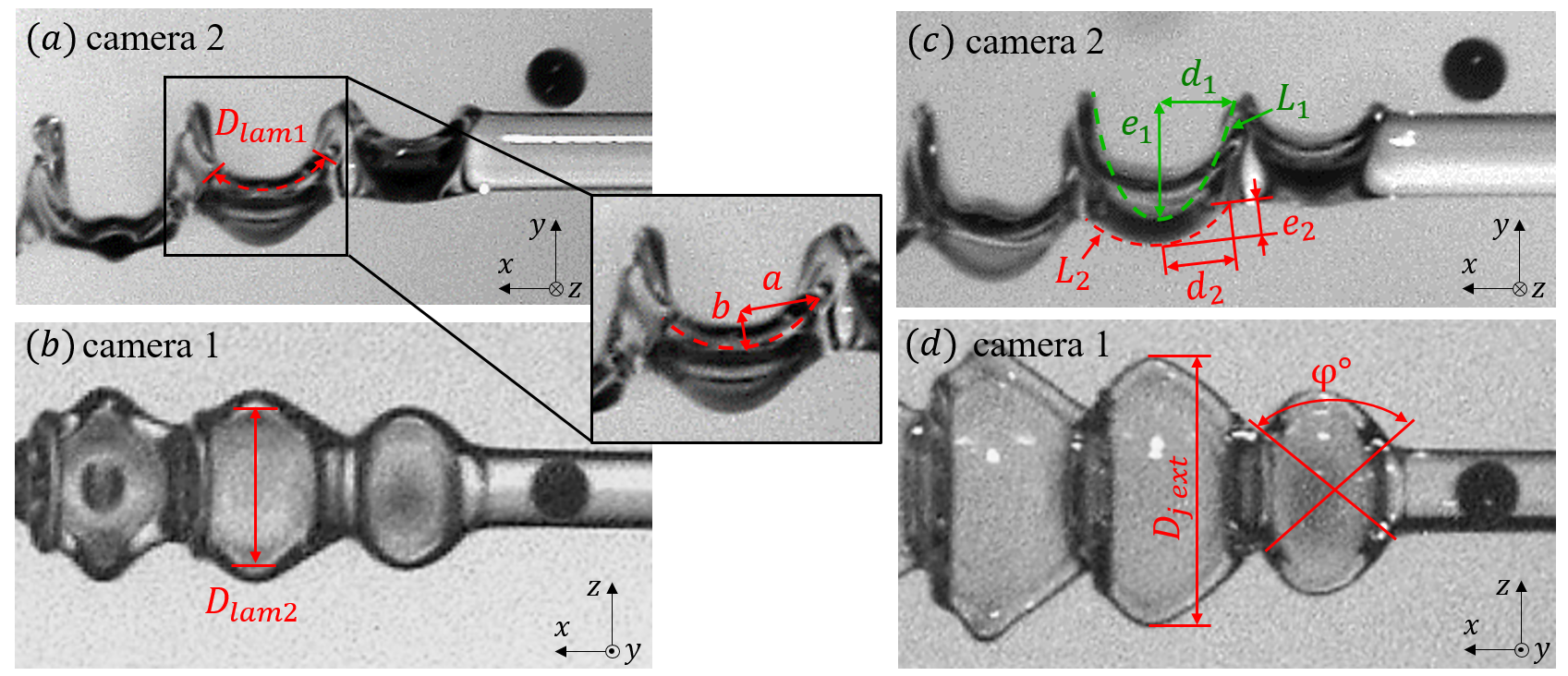}
  \caption{(a) front and (b) orthogonal view of a collision showing the parameters used to estimate the maximum surface extension reached by the drop (dark phase). (c) front and (d)  orthogonal view of a collision showing the parameters used to estimate the surface extension reached by the jet envelope (clear phase). The notations are explained in the text. }\label{fig:dmax}
\end{figure}

%To investigate the effects of liquid wettability on the collision outcomes, 
The surface extension of each immiscible phase is obtained via detailed image analysis using the public-domain software ImageJ (https://imagej.nih.gov/ij/). Collision pictures were taken with both front (camera 2) and orthogonal (camera 1) views, see figure \ref{fig:dmax}. The resolution of our imaging system is 10 $\mu m / px$ for general observations (regime determination, drop spacing and initial diameters), and 6 $\mu m / px$ for finer measurements (drop and jet deformation, drop fragmentation). Estimating the measurement uncertainty to $\pm$ 1 pixel leads to $\pm$ 5 \% and $\pm 3$ \% for the dimensions typically obtained with the general and finer resolutions, respectively. The temporal resolution of the image analysis is limited by the instants visible on the standing pictures. The time elapsed between successive images is $ 1/f_d$, causing a time resolution of 0.1 ms. To limit this sampling effect, movies are recorded using an aliasing frequency for the stroboscopic illumination. With this method, and except for the measurement of $t_{frag}$, 6 points per period are obtained, reducing the discretization to a period of 0.017 ms. Instants of maximal deformation are deduced from the temporal evolution of the quantity of interest, which is then fitted by a polynomial. The overall uncertainty of time obtained by this method is estimated to be less than 0.02 ms. Due to the difficulty in identifying accurately the drop fragmentation, the uncertainty in this measurement is in the range of $\pm 0.05 ms$.  

The extension of the drop is tracked following the dark phase (images (a) and (b)) and modelled  - from the first instants of the collision until it reaches its maximal extension  - by a bent disk whose unbent diameter is $D_{lam1} \approx D_{lam2}$. Here $D_{lam2}$ is directly measured from the orthogonal view (figure \ref{fig:dmax} (b)) whereas $D_{lam1}$ is  obtained from the front view (figure \ref{fig:dmax} (a))  as:

\begin{equation}
  D_{lam1}= \frac{\pi(a+b)}{2}\biggl(1+\frac{3\lambda^2}{10+\sqrt{4-3\lambda^2}}\biggr)
  \label{dmax_drop}
\end{equation}    
 The expression of $D_{lam1}$, for which $a$ and $b$ are measured on the image, corresponds to the Ramanujan approximation of an ellipse circumference; $\lambda= (a-b)/(a+b)$ is the ellipse eccentricity.  Note that by comparing $D_{lam1}$ and $D_{lam2}$ (data not shown, difference below 10\%) we verify that $D_{lam1} \approx D_{lam2}$. In reality, the thin lamella is surrounded by a toroidal rim which is ignored for simplicity. Instead, the surface of the lamella is  approximated by the one of a very flat cylinder whose diameter is $D_{lam} \approx D_{lam1} \approx D_{lam2}$ and height $h$, deduced from volume conservation using the not deformed drop. This approximation has been used and validated in several drop-drop collision studies \citep{Willis-Orme_2003, PlanchetteBrenn2017}. The  deformed drop surface thus reads: 
 
 \begin{equation}
\Sigma_{d\,lam}=\frac{ \pi D_{lam}^2}{2}\biggl(1+\frac{8}{3}\frac{D_d^3}{D_{lam}^3}\biggr)
\label{surface_dmax}
\end{equation}

where $D_{lam}$ is evaluated either by $D_{lam1}$ or $D_{lam2}$.

Quantifying the surface of one spatial period of the deformed jet  is difficult due to the complexity of the shape. To tackle this point, we estimate the surface as the sum of three contributions:
 \begin{itemize}
     \item$\Sigma_{1}= \pi L_1 D_{j\,ext} / 4$ is the area of the ellipse that forms on the impact side. The two axes of the ellipse are thus $L_1$ (see figure \ref{fig:dmax} (c), green dashed line) and $D_{j\,ext}$ (see figure \ref{fig:dmax} (d), red double arrow),   obtained using the  Ramanujan approximation with $e_1$, $d_1$ and   the orthogonal view, respectively. 
     \item $\Sigma_{2} = \pi L_2 D_{j\,ext} / 4$ is the area of the ellipse that forms opposite to the impact side. The two axes of the ellipse are thus $L_2$ (see figure \ref{fig:dmax} (c), red dashed line) and $D_{j\,ext}$,  obtained  using the  Ramanujan approximation with $e_2$, $d_2$ and   the orthogonal view, respectively.
     \item $\Sigma_{3}$ corresponds to a portion of the  side area connecting these two ellipses. We approximate the side area by the slant side of a truncated cone whose large and small radius are given by $R_{1}= \sqrt{\Sigma_1 / \pi}$ and $R_{2}= \sqrt{\Sigma_{2} / \pi}$, respectively. The height of the truncated cone is obtained by volume conservation $h_3=3\,(V_j+V_d)/[{\pi}(R_1^2+R_1R_2+R_2^2)]$ where $V_j=\pi {D_j}^2 L_j/4$ and $V_d=\pi {D_d}^3/6$ are the corresponding drop and jet volume. Only a portion of this slant area must be accounted for (see figure \ref{fig:dmax} (d), red continuous lines) which is equivalent to the angular fraction $\hat{\varphi}={2\varphi^\circ}/{360}$. Thus we obtain: $\Sigma_{3}=(R_1+R_2)\,\pi\,\sqrt{(R_1-R_2)^2+{h_3}^2}\,\hat{\varphi}$
     
 \end{itemize}
Finally, the surface extension of the jet is given by:
\begin{equation}
\Sigma_{j}= \frac{L_1}{2}\frac{D_{j\,ext}}{2}\pi+\frac{L_2}{2}\frac{D_{j\,ext}}{2}\pi+(R_1+R_2)\,\pi\,\sqrt{(R_1-R_2)^2+{h_3}^2}\,\hat{\varphi}
\label{surface_jmax}
\end{equation}

%\FloatBarrier
\section{Collision outcomes, classification and discussion}\label{res}

In this section we  first present the results obtained with the reference system, similar to the one of \cite{Planchette2018, PlanchetteBrenn2018}. The jet is made of silicone oil and totally wets the drops constituted of an aqueous solution of glycerol. 
%We then compare this result to the ones obtained using miscible liquids. After, we show how important the relative wettability of the two liquids is by using for the drops a liquid that totally wets the one of the jet. Finally, we quantify the effect of the spreading parameter onto the transitions between various regimes.

\subsection{Immiscible, totally wetting liquid jet}\label{sec:G5_SOM5}

\begin{figure}
  \centering 
  \includegraphics[width=\textwidth]{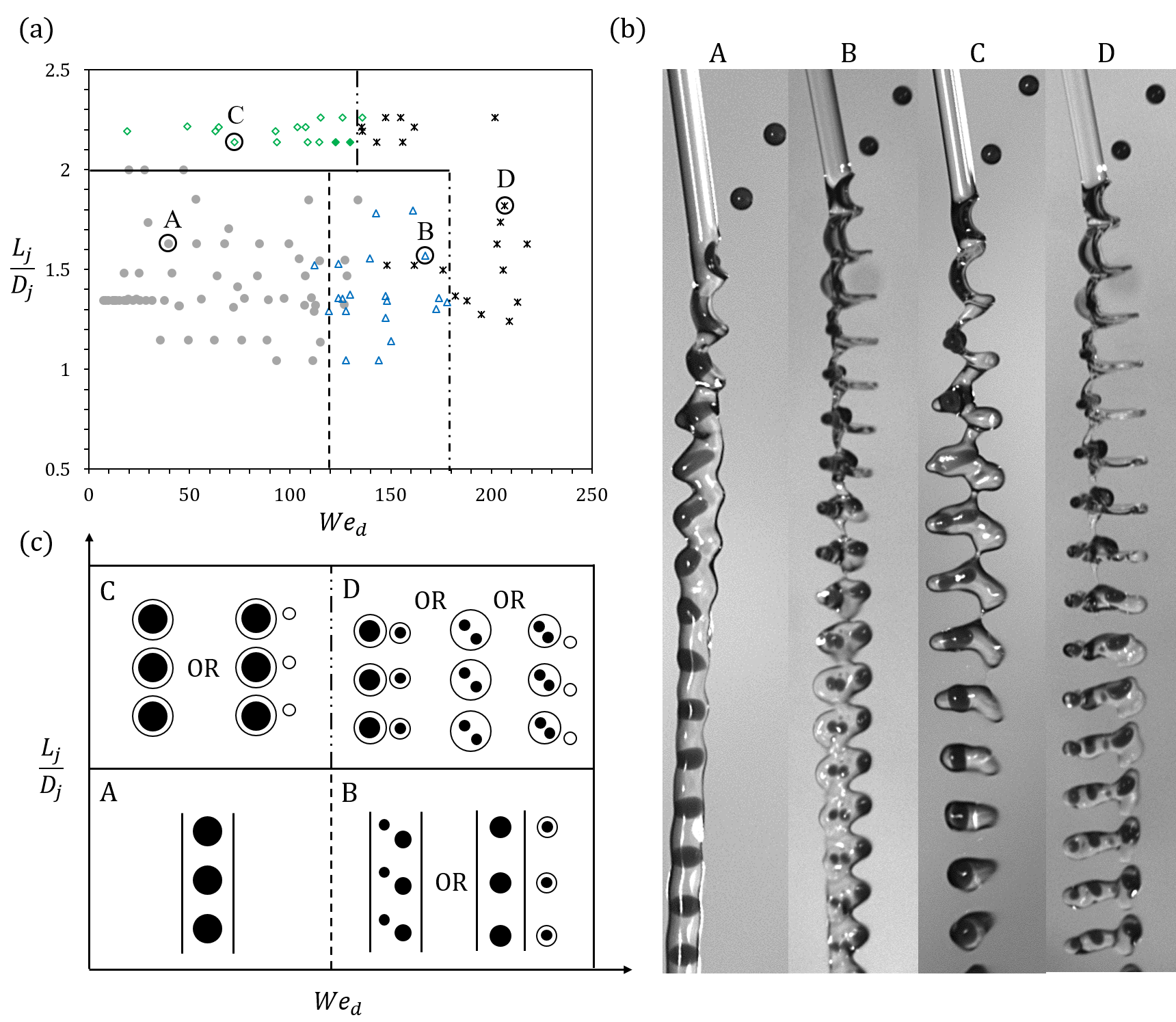}
  \caption{(a) Experimental regime map for drops of G5 and a jet of SO 5. \textit{Drops in jet} grey full circles; \textit{fragmented drops in jet} empty blue triangles; \textit{encapsulated drops} green diamonds (empty without satellite, full with satellites); and \textit{mixed fragmentation} black crosses. Black empty circles correspond to the images of part (b), the lines are guides for the eye. (b) Images illustrating the four regimes with following collision parameters: (A) ${L_j}/{D_j}=1.63$, $We_d=39.42$; (B) ${L_j}/{D_j}=1.57$, $We_d=167.05$; (C) ${L_j}/{D_j}=2.14$, $We_d=72.54$; and (D)  ${L_j}/{D_j}=1.82$, $We_d=206.55$. (c) Sketches of the observed regimes; black: drop liquid, white: jet liquid.}
\label{fig:G50_SOM5}
\end{figure}

The collision outcomes we observe using a continuous jet of SO 5 with a stream of G5 drops correspond to the four regimes described in \cite{PlanchetteBrenn2018} and recalled below for completeness. The occurrence of these regimes is described in figure \ref{fig:G50_SOM5}(a) using the geometric parameter $L_j/D_j$  and the drop Weber number $We_d$. $L_j/D_j$ was introduced in \cite{PlanchetteBrenn2018} to predict the capillary fragmentation threshold. In the same study, a modified Weber number was used to describe the inertial fragmentation limit. The latter being found unsatisfactory when applied to liquids with different wetting and miscibility (data not shown), it is replaced by $We_d$. \\
%found to work as a modified Rayleigh criterion enabling to distinguish continuous and capillary fragmented jet \citep{Planchette2018, PlanchetteBrenn2018}. A second parameter corresponding to a modified Weber number was introduced by \cite{ PlanchetteBrenn2018} and used to describe an inertial fragmentation mechanism. Yet, this parameter being unable to bring together the transitions obtained in this study with partial and total wetting (data not shown), we have decided not to use it here. Instead we present our result using the drop Weber number $We_d= \rho_d D_d U^2/\sigma_d$ calculated using the relative velocity $U$. This choice will be discussed later.\\ 

\begin{itemize}
\item \textit{drops in jet} occur when the jet immediately engulfs the impacting drop forming a continuous jet with a regular and periodic distribution of embedded drops, see  figures \ref{fig:G50_SOM5}(b) and (c)-A for a picture and sketch, respectively. This regime can be found for low to moderate $We_d$ ($\leq 120)$  and for a ratio ${L_j}/{D_j}$ lower than a critical value close to 2, see figure \ref{fig:G50_SOM5}(a), full grey circles.\\ %Due to the fact, that this regime is of great interest for future advanced fibre production, the geometric parameters after collision are important. This includes the diameter of the drop/jet stream $D_{dij}$ after collision as well as the spatial period of the embedded drops $L_{dij}$. A cylindrical continuous jet, neglecting viscous losses and volume as well as momentum conversation lead to following equations:
%\begin{equation}
%D_{dij}=\tilde{D}_{d} \frac{{\left(1+ m_j  /m_d\right)}^{1/2}{\left( 1+ \rho_d m_j  / \rho_j m_d \right)}^{1/2}}{\left[{{\left(\cos{\alpha}+ m_j u_j/ m_d u_d \right)}^2+{{\sin}^2 \alpha }}\right]^{1/4} }
%\label{D_dij}
%\end{equation} 
%\begin{equation}
%\frac{L_{dij}}{L_d}=\frac{\sqrt{{\left(\cos{\alpha}+ m_j u_j/ m_d u_d \right)}^2+{{\sin }^2 %\alpha }} }{1+ m_j  /m_d}
%\label{L_dij}  
%\end{equation}
%$\tilde{D_d}$ represents an equivalent drop stream diameter, which is defined as $\tilde{D_d}=(2/3)(D_d^3/L_d)$. $m_j$ and $m_d$ quantify the mass of one jet portion and one corresponding drop. They are given by: $m_j=\pi/4D_j^2L_j\rho_j$ and $m_d=\pi/6D_d^3\rho_d$ \citep{PlanchetteBrenn2018}.

\item \textit{fragmented drops in jet} correspond to drops that are totally encapsulated by the jet in which they fragment, leaving the jet continuous, see picture  \ref{fig:G50_SOM5}(b)-B. In most cases, all drop fragments remain inside the continuous jet. In some cases, however, some of them are expelled from the jet, dragging a thin layer of jet liquid, but  leaving the main part of the jet continuous,  see the sketches of figure \ref{fig:G50_SOM5}(c)-B. This regime can be found for higher $We_d$ in the range of $120\leq We_d \leq180$ and for ${L_j}/{D_j}\leq2$, (see  figure \ref{fig:G50_SOM5}(a), empty blue triangles).\\ % can be found for higher $We_d$ in the range between $120\leq We_d \leq180$ and for ${L_j}/{D_j}\leq2$, e.g. below the capillary fragmentation threshold of the jet (see  figure \ref{fig:G50_SOM5}(c), empty blue triangles ). After the point of impact the initial kinetic energy of the drop is mainly converted into viscous losses and  surface energy. The increasing surface energy consequently leads to an increasing surface area of the drop which may then fragment as illustrated by the picture of figure \ref{fig:G50_SOM5}(b)-B. In most cases the fragmented drops remain inside the continuous jet. Once the jet recovers its cylindrical shape, the drop fragments are found to follow a spatial periodicity. In some cases, however, most likely if the kinetic energy of the drop further increases, parts of the drop fragments are expelled from the jet. These small fragments drag a thin layer made  of the jet liquid, but  the main part of the jet remains continuous,  see the sketches of figure \ref{fig:G50_SOM5}(c)-B.\\     
\item \textit{encapsulated drops} correspond to a regular stream of capsules, each of them being made of exactly one not fragmented aqueous drop (core) coated by a layer of jet liquid (shell), see picture and sketches of figure \ref{fig:G50_SOM5}(b) and (c)-C. The presence  or absence  of satellite drops made of the jet liquid only can be observed  \citep{Planchette2018} and is reported in figure  \ref{fig:G50_SOM5}(a) in the form of filled or empty diamonds, respectively. This regime is found for ${L_j}/{D_j} > 2 $ and moderate values of $We_d$ ($\leq135$).\\
%are observed for ${L_j}/{D_j}$ larger  than the critical value of 2 which corresponds to the capillary fragmentation limit of the jet. For the drops not to fragment, moderate values for $We_d$ ($\leq135$) must be observed, figure \ref{fig:G50_SOM5}(a)-C. Thus we obtain  a continuous stream of capsules made of drops encapsulated by the jet liquid. The fragmentation of the jet takes place in the thinner portion found between two consecutive impact points, see picture and sketch of figure \ref{fig:G50_SOM5}(b) and (c)-C. It is important to mention, that two sub-regimes can be distinguished \citep{Planchette2018} based on the presence or absence of satellite drops made of the jet liquid only. In figure \ref{fig:G50_SOM5}(a), empty diamonds correspond to the absence of satellites as illustrated by the picture  of figure \ref{fig:G50_SOM5}(b)-C, while filled diamonds correspond to the presence of satellite drops. Encapsulated drops with satellites are not shown in this paper.\\           
\item \textit{mixed fragmentation}  is observed  when both the drop and jet fragment and is illustrated in figure \ref{fig:G50_SOM5}(b) and (c)-D. It is found for high $We_d$ and for any value of $L_j/D_j$, see figure \ref{fig:G50_SOM5}(a)-D (black stars).\\  
\end{itemize}

%%here
The regime occurrence, presented in the form of the map of figure \ref{fig:G50_SOM5}(a), summarizes the results obtained with a jet of SO 5 and drops of G5.
It confirms that the dimensionless parameter $L_j/D_j$ is well suited to predict the capillary fragmentation of the jet. The critical value of approximately 2 deviates from the theoretical value of $\pi$ found by Rayleigh. This can be explained by the effects of inertia and the presence of two phases. %, where the encapsulated drops have a defined Laplace pressure. 
Both may affect the flow in the jet and thus its stability towards periodic perturbations.
The second dimensionless parameter chosen for this study is $We_d$. Accounting for the uncertainty on the measurements of drop diameters (typically $\pm{5}$ \%), $We_d$  appears appropriate to distinguish other fragmentations from inertial origins. For the domain $L_j/D_j<2$, a first fragmentation concerning the drops is found for a critical value of approximately $120$, followed by a second one for which both the drop and jet fragment around $180$. %Note that the two points of mixed fragmentation found for $L_j/D_j\approx 1.5$ and $We_d \approx 150-170$ are most likely 
For the domain where the jet is unstable ($L_j/D_j>2$), the inertial fragmentation is found around $We_d \approx 135$ and leads to the regime of mixed fragmentation.

This regime description and map obtained for immiscible and totally wetting jet liquid raise several important questions. First, how universal are the observed regimes? Especially, can they be obtained using any immiscible liquids or must  certain wetting conditions be fulfilled? Further, how the previous regime description must be modified to account for miscible liquids? For liquid couples producing the same kinds of regimes, how universal are the observed fragmentation mechanisms and thresholds? %Are the proposed parameters $L_j/D_j$ and $We_d$ still valid? Do the critical values remain the same? 

To answer these questions, we have systematically varied the combination of liquids and start in the next section with the extreme case of miscible liquids.
%, enabling to address each point separately. Let us start with the extreme case where liquids are miscible.
\FloatBarrier
\subsection{Miscible liquids}\label{misc}
Here the drops are still made of G5, but the silicone oil has been replaced by an aqueous solution of ethanol and glycerol (EtOH).

As mentioned, the combination of miscible liquids is of great interest in view of utilizing the drop-jet collisions to achieve encapsulation, especially for  medical and biomedical applications \citep{Visser2018, Kamperman2018, fibers}. In this context, the main question concerns the  final liquid distribution. 

\begin{figure}
  \centering 
  \includegraphics[width=\textwidth]{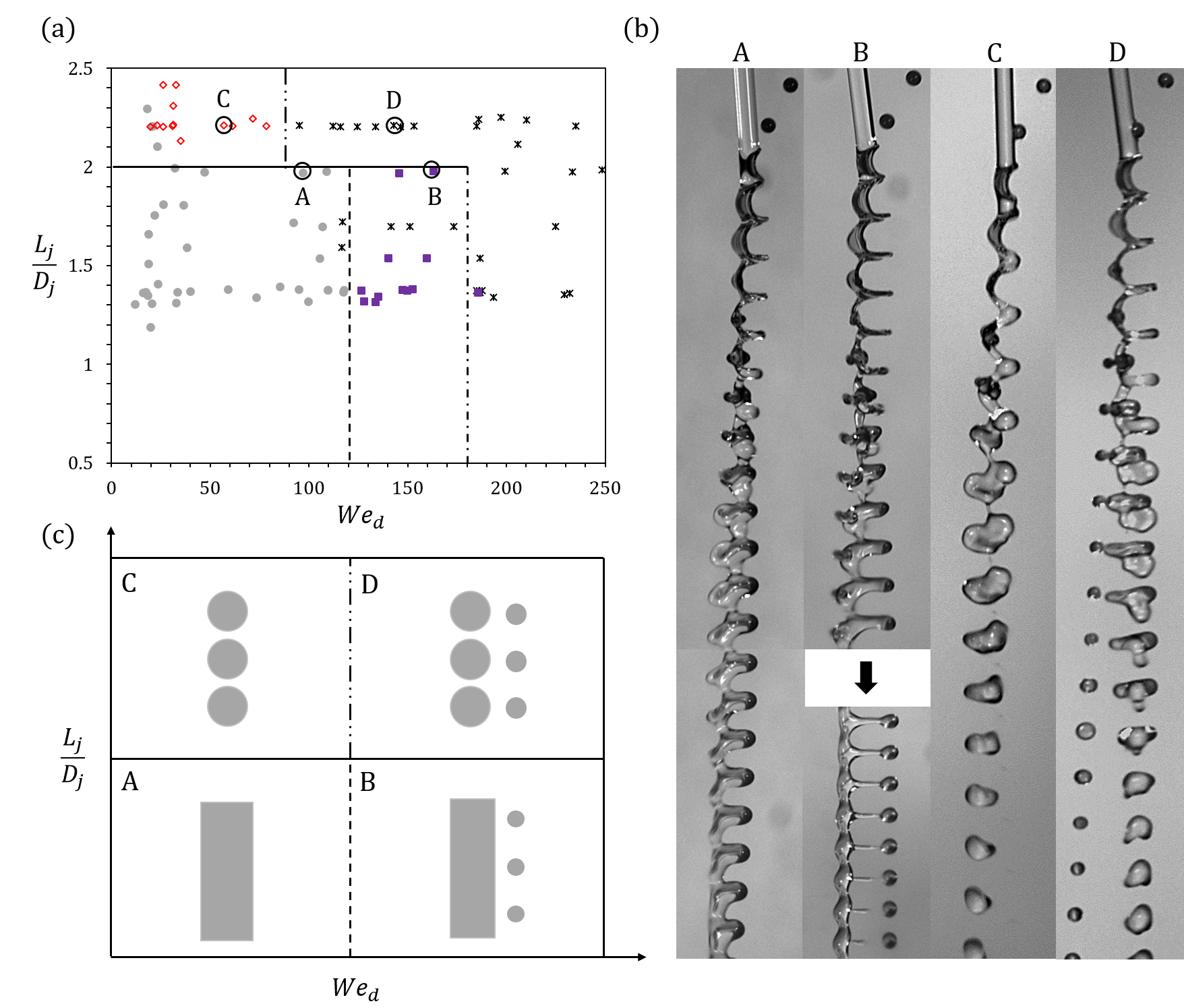}
  \caption{(a) Experimental regime map obtained with glycerol G5 as droplet and EtOH as jet.
  \textit{Drops in jet} grey full circles; \textit{fragmented drops and continuous jet} full purple diamonds; \textit{simple fragmented jet}  red diamonds; and \textit{mixed fragmentation} black crosses. Black empty circles correspond to the images of part (b), the lines are guides for the eye. (b) Images illustrating the 4 regimes with following collision parameters: (A) ${L_j}/{D_j}=1.97$, $We_d=47.0$; (B) ${L_j}/{D_j}=1.97$, $We_d=163.22$; (C) ${L_j}/{D_j}=2.21$, $We_d=56.96$; and (D) ${L_j}/{D_j}=2.2$, $We_d=146.25$. The black arrow indicates a discontinuity in the image.
  (c) Schematic representation of the observed regimes; black: drop liquid, white: jet liquid, grey mixture of both liquids. }
\label{fig:G50_EtOH}
\end{figure}

Since no more interface between the jet and drop liquid exists, the regime classification must be redefined. We propose to replace the fragmentation or not of each phase by the presence or not of each liquid in each entity (in the same as initially or in separated ones). Following this approach, we obtain the regimes described below and illustrated in figure \ref{fig:G50_EtOH}; with (a) a regime map representing their occurrence, (b)  pictures and (c) sketches.\\

\begin{itemize}
\item   \textit{drops in jet}  is still  observed with miscible liquids.  We have decided to keep the wording despite the absence of an interface since the liquid of the drops remains very localized in the jet, similarly to the case of immiscible liquids, see figures \ref{fig:G50_EtOH} and  \ref{fig:G50_SOM5} (b)-A. 

This observation is in agreement with the comparison of $<x_{diff}>$, the typical diffusion length during the process duration, and the drop diameter of $200$ \textmu m. Indeed, estimating the process duration by  $t_{d\,osc}= \sqrt{{\rho_dD_d^3}/{\sigma_d}}$, a fraction of the drop oscillation period  (justified later), we find   $<x_{diff}> = \sqrt{2D t_{d\,osc}} \approx 1$ \textmu m \citep{Einstein1956, Christen2010}.

A direct consequence of this finding is that even with low viscosities ($ \approx 5$ mPa.s), it is possible to produce a continuous jet with a regular composition pattern. 
 For comparably fast solidification processes, 
 fibres could be obtained with regular and controlled distribution of some compounds in a main cylindrical body showing great potential for tissue engineering \citep{Khademhosseini2480}.  
 
Interestingly, both the inertial and capillary limits of this regime seem similar to the ones found for immiscible liquids, as indicated by  the dashed and continuous lines of figure \ref{fig:sigma_jmax}(a), found  at  $We_d \approx 120$,  and $L_j/D_j\approx2$, respectively.
 \\
\item  Above the first inertial fragmentation limit, for $L_j / D_j<2$ and $We_d >120$, a  continuous jet containing part of the drop liquid plus a stream of satellite drops made of both the drop and jet liquids can be seen,  figure \ref{fig:G50_EtOH}(a) purple squares. The composition of the jet and satellite drops is revealed by the dye distribution. We call this regime \textit{fragmented drops and continuous jet}.  Illustrations, picture and sketch, are provided  in figures \ref{fig:G50_EtOH}(b-c)-B. It is important to mention that the fragmentation of the droplets inside the jet does not exist, since there is no interface. Thus, what we observe here under the term \textit{fragmented drops and continuous jet} corresponds only to a fraction of what is classified  as  \textit{fragmented drops in jet} with immiscible liquids.  \\

\item $\,$Above the capillary limit (${L_j}/{D_j}>2$) and for low to moderate values of $We_d$, the jet fragments, figure \ref{fig:G50_EtOH}(a), red diamonds.
Except for the absence of an interface inside the fragments, this regime called \textit{simple fragmented jet}, is similar to the \textit{encapsulated drops} observed with immiscible liquids in the absence of satellite drops, see figure \ref{fig:G50_EtOH}(b-c)-C. \\   

\item For $L_j/D_j>2$ and  for  $We_d>90$ we observe that  the jet fragments and that the drop liquid is found in more entities after than before the collision (see figure \ref{fig:G50_EtOH}, (a)). By analogy with immiscible liquids, we refer to it as \textit{mixed fragmentation}. Note, however, that due to the absence of interface,  the liquid distribution obtained with miscible liquids corresponds only to a fraction of the ones observed under the same wording using immiscible liquids, see figures \ref{fig:G50_SOM5} and \ref{fig:G50_EtOH}, (b-c)-D.\\   
\end{itemize}

By replacing the criteria based on topology changes (fragmentation) of each immiscible phase by criteria based on the spatial distribution of the liquids (especially drop liquid present in more entities than initially), the description of the collision outcomes obtained with miscible liquids appears very similar to that found with immiscible liquids.
Four main regimes separated by two fragmentation mechanisms of capillary and inertial origins are obtained in both cases.

The  transitions of inertial origin found for $L_j/D_j<2$ are very similar to the one observed with immiscible liquids. It is worth noting that, despite the new definitions, the  regime occurrence  remains mostly unchanged within the experimental uncertainty. To tackle the lack of interface between the two liquids, the drop is considered as fragmented only if its liquid is found in more entities after than before the collision. Thus, the liquid distribution corresponding to \textit{fragmented drops in jet} with immiscible liquids may be re-formulated for miscible liquids either as \textit{fragmented drops and continuous jet}, or as \textit{drops in jet}. Surprisingly,  this neither leads to a  reduced occurrence of \textit{fragmented drops and continuous jet}, nor to  a  widened observation of \textit{drops in jet}. Indeed, comparing the results obtained for  $L_j/D_j<2$ and $120<We_d<180$, the split between the drop fragments and the continuous jet  is as frequently observed with miscible liquids (80 \% or 12 out of 15 points, see figure \ref{fig:G50_EtOH}(a)) as with  immiscible ones (72 \% or 18 out of 25 points). Furthermore, no significant shifts in these two transitions are measured, which is far from being trivial.

Different complex and possibly coupled, at least co-existing, effects are expected to influence the fragmentation limits: the presence of Marangoni flows, the absence of Laplace pressure maxima caused by the drops in the jet,  the absence of thin oil films which  concentrate most of the viscous dissipation \citep{Planchette2012}, or more generally different energy transfer as observed for drop impacts onto liquid films for which liquid miscibility suppresses crown formation \citep{Chen_2017}. The consequences of each separate effect on the fragmentation limits remain unclear. All in all, it seems that the differences observed in these two types of collisions compensate to produce the same transitions below $L_j/D_j<2$, which is quite remarkable. Additionally, one cannot exclude that, on the typical time scale of the collisions (1 ms), different liquids suddenly brought into contact, may see the presence of a diffuse interface reminiscent from their composition difference, and so even if they are miscible \citep{shikhmurzaev}.

The inertial limit found for $L_j/D_j>2$ is observed for  $We_d \approx 90$ instead of $130$ for immiscible liquids.  A first possible explanation  to this shift could be the change of criteria used to define the collision outcomes. The liquid distribution corresponding with immiscible liquids to \textit{encapsulated drops with satellites} may appear as \textit{mixed fragmentation }with miscible liquids. Indeed, due to the liquid miscibility, it is not possible to exclude minutes of drop liquid in the satellites. Yet, even if the \textit{encapsulated drops with satellites} were to be classified with the \textit{mixed fragmentation}  (see figure \ref{fig:G50_SOM5}a), the transition would be found for $We_d$ equal to $120$, far above $90$.

Regarding the capillary fragmentation limit, we  note the presence of three points  at low $We_d$, where the jet is stable despite  ${L_j}/{D_j}>2 $ (values ranging from 2.1 to 2.3). We attribute this deviation to possible Marangoni flows from the undisturbed jet sections (low surface tension) to the points of drop impact (high surface tension). These flows could contribute in restoring the cylindrical shape of the jet and therefore stabilize it.

\FloatBarrier
\subsection{Immiscible, totally wetting  drop liquid}\label{total}

While the effects of liquid miscibility appear to be very limited on the collision outcomes and their occurrence, one could ask about the role of liquid wettability. Indeed, the reference system was designed - using a totally wetting jet liquid - to promote the encapsulation of the drops by the jet. Thus, several questions  relevant for encapsulation applications arise: can drops be engulfed  in an immiscible jet against the wetting thermodynamics? Is there a critical level of inertia to be reached? Do the drops fragment and get only partially encapsulated?  
Much has been done about encapsulation and its limits in the field of microfluidics. Classically, emulsions (simple or multiple) are obtained by drop formation caused by the break-up of a disperse phase in a continuous one  (at one position or in subsequent ones, respectively) \citep{Chu_2007}. Since the drop liquid is injected into the continuous phase, the effect of relative wettability between the liquids to emulsion stability is rather limited. As an alternative to the classical microfluidics approach, it was proposed to bring droplets of different liquids, dispersed in the same continuous phase, into  contact to produce the encapsulation of one drop by the other \citep{Deng2013}. Controversial conclusions were drawn, attributing the drop engulfment  to a difference of Laplace pressure. Indeed,   the thermodynamics of wetting should control the final  distribution of the two drop liquids \citep{comment_deng_2013}. Experiments show deviations from the pure thermodynamic configuration. Drops were engulfed by each other, despite partial wetting of the two liquids. These findings were attributed to dynamic effects, and  especially to the flow of the continuous phase \citep{reply_comment_deng_2013}. % Yet the flow was laminar and other effects such as gradients of surfactants were not considered. In any case, this study shows that ability  were obtained.... 
Similar dynamic engulfment was reported for drop-drop collisions in air using n-hexadecane and water \citep{Wang_2004} and predicted by recent numerical simulation \citep{semprebon_2018}.  Yet, only  partial wetting was considered, and none of the studied geometries was close to the one of drop-jet collisions. Thus, the questions remain, and we are not aware of any investigation of this kind for in-air microfluidics. While the case of partial wetting will be studied in the next section,  we focus here on the extreme case for which the drop liquid totally wets the jet opposing their engulfment. 

The droplets consist of SO 5 coloured with blue dye, whereas the jet consists of G5,  which is kept transparent.%NEWLY REMOVED As mentioned before, and in agreement with the surface and interfacial tensions reported in table \ref{tab:fluidproperties}, upon contact between the two liquids, the silicone oil totally spreads at the surface of the glycerol solution.\\
\begin{figure}
  \centering 
  \includegraphics[width=\textwidth]{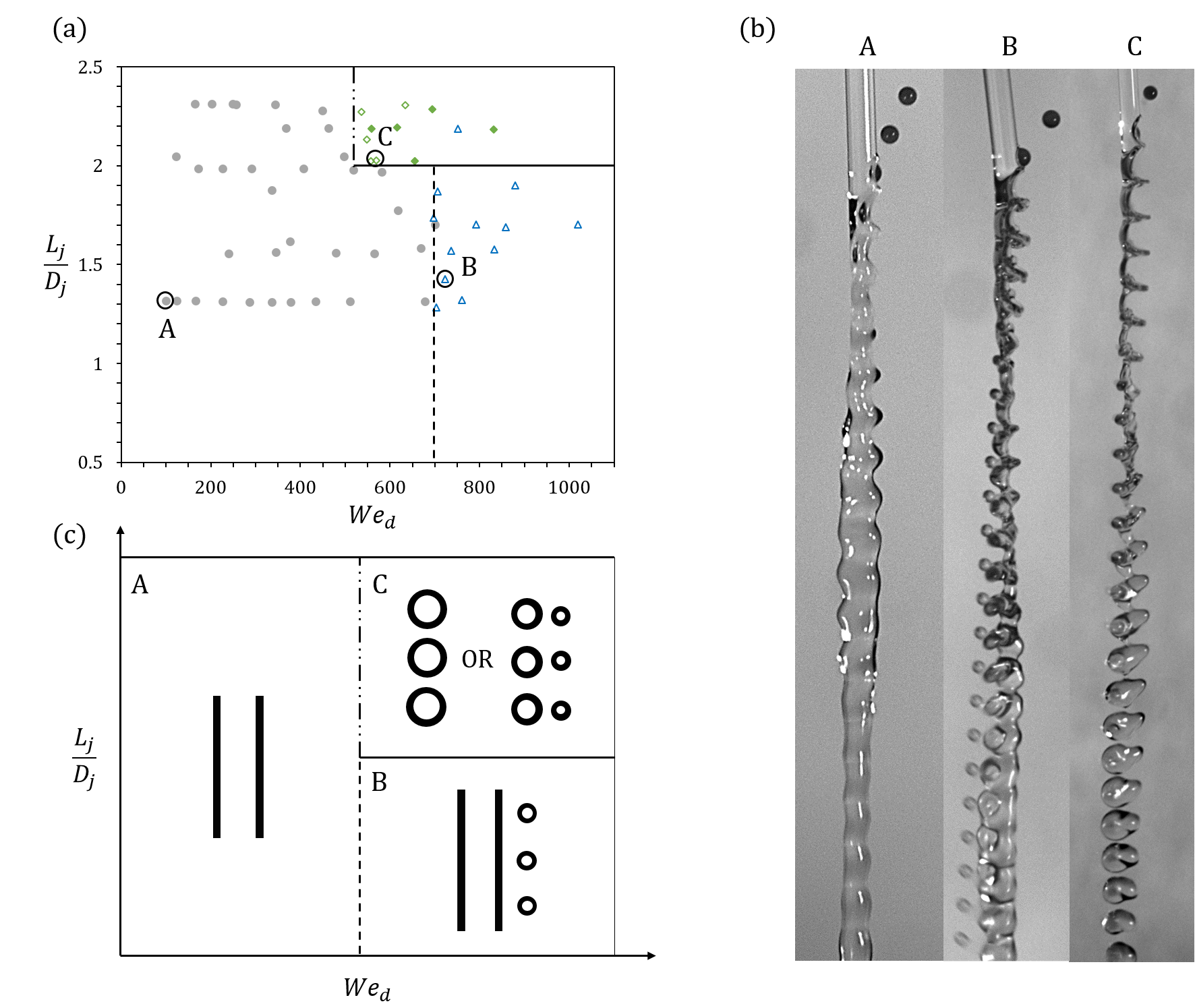}
  \caption{(a) Experimental  regime map for collisions of drops of SO 5 and a jet of G5.  (A) \textit{coated jet} full grey circles; (B) \textit{coated jet with satellites} empty blue triangles, (C) \textit{fragmented coated jet} green diamonds (empty without satellites, full with satellites). The lines are guides for the eye and the black circles correspond to the pictures shown in the part (b). (b) Pictures illustrating the 3 regimes and corresponding to (A)
  ${L_j}/{D_j}=1.31$, $We_d=100.14$ ; (B) ${L_j}/{D_j}=1.28$, $We_d=702.71$; (C)  ${L_j}/{D_j}=2.02$, $We_d=568.94$. (c) Schematic representation of the observed regimes; dark: drop liquid (SO 5), white : jet liquids (G5).}
\label{fig:SOM5_G50}
\end{figure}

The conducted experiments cover the same range of parameters as in the first two series  ($1.0<L_j/D_j<2.2$; $2<U<10$ m/s ). The large values of $We_d$ ($100<We_d<1000$) are caused by the silicone oil properties. Qualitatively, basing our classification on the fragmentation of each phase, only three  regimes can be distinguished. As shown by the pictures in figure \ref{fig:SOM5_G50}(b), the liquid spatial distribution of these  regimes is completely different from those found in other experiments (section  \ref{sec:G5_SOM5} and \ref{misc}). More precisely, we obtain:
\FloatBarrier

\begin{itemize}
\item$\,$ A \textit{coated jet}. The jet does not fragment and is quickly recovered by a film made of the impacting drop liquid, see figure \ref{fig:SOM5_G50}(b) and (c)-A. Note that neither the whole droplets nor part of them can be engulfed into the jet. 
It is the most frequently observed regime, and its occurrence extends beyond the \textit{drops in jet} observed with the previous systems,  see figure \ref{fig:SOM5_G50}(a), full grey circles. \\

\item$\,$ A \textit{coated jet with satellites} which corresponds to a continuous jet accompanied by a stream of small satellite droplets. Both the jet and the satellite droplets are made of a core of G5 coated by SO 5, the drop liquid, see figures \ref{fig:SOM5_G50}(b-c)-B.  
This regime is observed  for $We_d\geq700$ ($U\geq8$ m/s) and for ${L_j}/{D_j}\leq2$, see figure \ref{fig:SOM5_G50}(a), empty blue triangles. For ${L_j}/{D_j}\leq2$ and within the velocity range screened in these experiments (e.g. up to $U\approx 10$ m/s) it is not possible to break up the jet. Finally, note that this regime corresponds to the fragmentation of the drops, while the jet remains continuous, making it analogous to \textit{fragmented drops in jet} (section \ref{res}) or to \textit{fragmented drop and continuous jet} (section \ref{misc}), apart of course from a very different spatial distribution of the liquids. \\   
\item$\,$A \textit{fragmented coated jet}. The  jet fragments into a regular stream of droplets, all of them being coated by a thin film of silicone oil, see figure \ref{fig:SOM5_G50}(b-c)-C.  The resulting number of  fragments may be equal to the initial number of drops, similarly to what is observed for \textit{encapsulated drops without satellites}, empty diamonds in figure \ref{fig:SOM5_G50}(a). Sometimes, the fragment number is a multiple of the initial drop number, which  is indicated by filled diamonds. The regime is observed for large $L_j/D_j$ and large $We_d$.\\ 
\end{itemize}

The absence of a capillary limit for low $We_d$ at $L_j/D_j \approx 2$ is remarkable, but can be explained by different factors. First, one cannot exclude that such a limit exists for a higher critical value. Furthermore, here, as in the rest of the paper, we observe the jet only over a finite distance classifying it as non-fragmenting if it recovers a quasi-cylindrical shape. The subsequent evolution found for larger timescales is not considered. 
Finally,  this absence may be caused by insufficient inertia or, said differently, by ineffective disturbances. Indeed, the drop spreading around the jet is expected to be very dissipative \citep{PlanchetteBrenn2018, Planchette2012, Chen_2017}, and the fragmentation limit at $L_j/D_j\approx 2$ is recovered  for $We_d>500$. 

Another remarkable finding is the impossibility to force, even partially, the drops into the jet. In the present configuration, the capillary  spreading driven by $S>0$ is facilitated by the ratio of Laplace pressures between the drops and jet:
$\Delta{p_{d\,SO\,5}}/{\Delta{p_{j\,G5}}}={2 D_j \sigma_d}/({D_d \sigma_j})\approx0.85$
in contrast to the reversed situation, for which we find: 
$\Delta{p_{d\,G5}}/{\Delta{p_{j\,SO\,5}}}\approx 10$. A further increase of the relative velocity causes the fragmentation of the drops and leaves  a stream of coated satellites and a coated jet. Keeping the same liquid combination, the diameter ratio $D_j/D_d$ (here $\approx1.5$) should be equal to $17$ in order to reach the Laplace pressure ratio of the reversed configuration. This would considerably limit the encapsulation capacity of the jet and, at
this stage, the use of  surfactants appears more promising  to tackle this
challenge.

\subsection{Immiscible, partially wetting jet liquid}\label{partial}

Having revealed that total but reverse wetting prevents encapsulation of the drops, it is legitimate to ask if partial wetting can overcome this limitation.  Indeed, while the addition of surfactants could suppress unfavorable total wetting, it cannot reverse it, but rather produces partial wetting conditions.  Here partial wetting is obtained using drops of the aqueous glycerol solution G5  (coloured with blue dye) with a liquid jet made of n-hexadecane (transparent).  The geometric parameter ${L_j}/{D_j}$ and the drop inertia ($We_d$) are varied  in the same ranges as in the other series of experiments. Figure \ref{fig:G50_Hexa}(a) illustrates the regime map of the collision outcomes, the associated pictures are shown in (b) and sketches of the liquid distribution are provided in (c). In short, using n-hexadecane for the jet, we observe the same regimes as with silicone oil (or perfluorodecalin). More precisely, we have:

\begin{figure}
  \centering 
  \includegraphics[width=\textwidth]{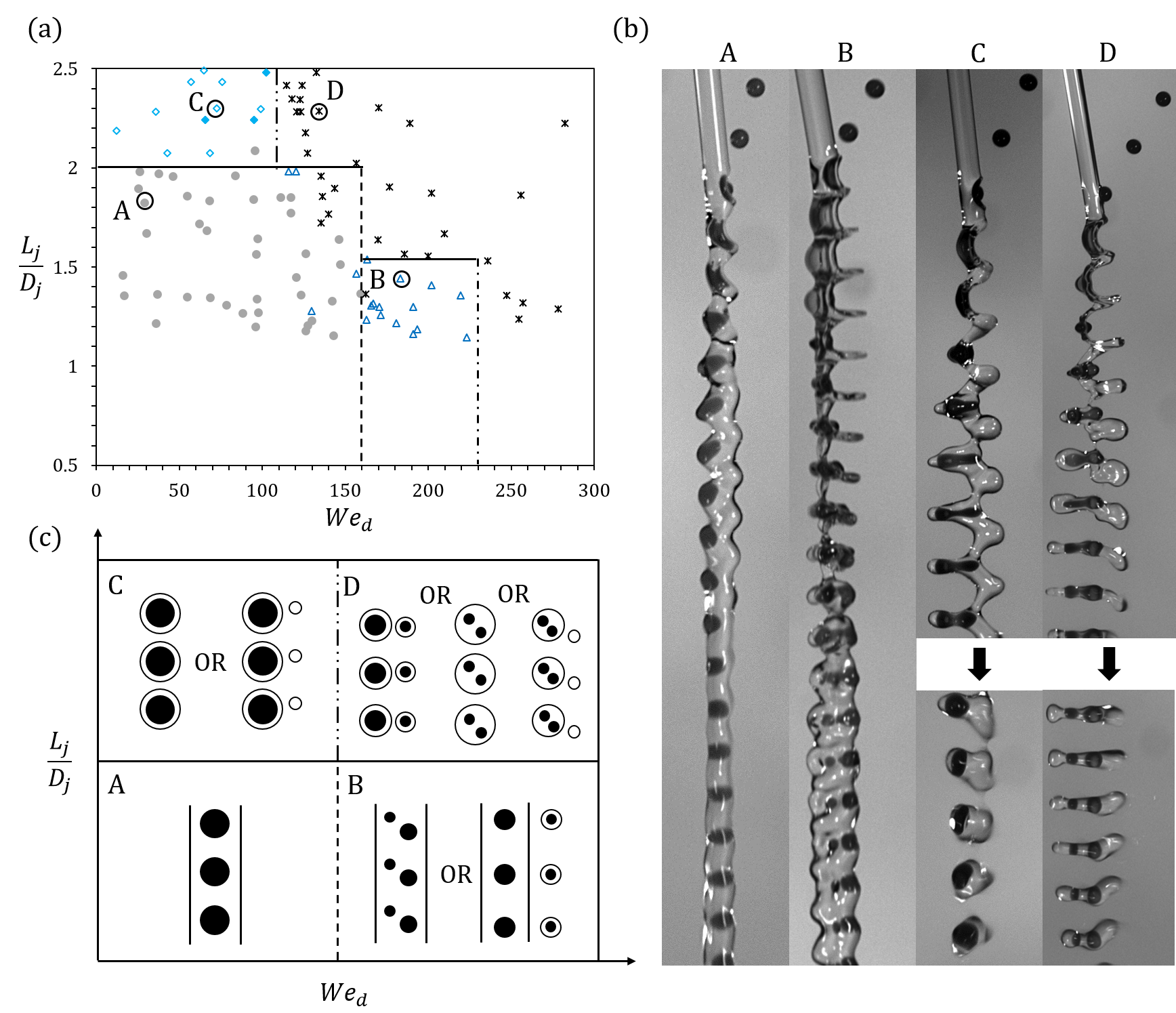}
  \caption{(a) Experimental regime map for drops of G5 and a jet of n-hexadecane. (A) \textit{drops in jet} full gray circles,  (B) \textit{fragmented drops in jet} empty blue triangles, (C) \textit{encapsulated drops} blue diamonds (empty without satellites, full with satellites),  and (D) \textit{mixed fragmentation} black crosses. The lines are guides for the eye, the black circles correspond to the pictures of part (b). (b) Collisions pictures corresponding to (A)  ${L_j}/{D_j}=1.82$, $We_d=28.87$; (B) ${L_j}/{D_j}=1.44$, $We_d=183.07$; (C) ${L_j}/{D_j}=2.29$, $We_d=72.62$; and (D) ${L_j}/{D_j}=2.28$, $We_d=134.21$. (c) Schematic representation of the observed regimes. Black: drop liquid, white. jet liquid.}
\label{fig:G50_Hexa}
\end{figure}

\begin{itemize}
\item$\,$\textit{drops in jet} for low to moderate kinetic energy  and for  ${L/_j}{D_j}<2$. It is worth mentioning that even for very low kinetic energy  (down to $We_d \approx 15$, the minimum reached in this study),  the drop is always fully encapsulated by the jet liquid. We do not observe the "sticking" regime found by \cite{Wang_2004} and predicted by \cite{semprebon_2018}, both for drop-drop collisions. \cite{Wang_2004} observed \textit{adhesive merging} in case of head-on drop-drop collisions with pure water and n-hexadecane droplets of equal diameter ($\approx300$ \textmu m) at a Weber number of $\approx4$ related to the liquid properties of pure water, which is lower than the lower value of $We_d$ reached here.\\

\item$\,$ \textit{Fragmented drops in jet} observed in a rather smaller area of the regime map ($170<We_d<230$ and ${L_j}/{D_j}<1.5$) when compared to the favorable total wetting configuration ($120<We_d<180$ and ${L_j}/{D_j}<2$). Beyond these limits,  it is replaced by \textit{mixed fragmentation}. 
\item$\,$ \textit{Encapsulated drops}  found for small and moderate $We_d$ and above ${L_j}/{D_j} \approx 2$ with and without satellite droplets as for a jet of silicone oil.
\item$\,$ \textit{Mixed fragmentation},  observed above $We_d = 110$ for ${L_j}/{D_j>2}$, with a threshold level which increases  for $1.5<{L_j}/{D_j<2}$ to $We_d\geq130$, finishing at $We_d\geq240$ for ${L_j}/{D_j}\leq1.5$.

\end{itemize}

The results obtained with n-hexadecane  show the same four regimes as with silicone oil. The existence of two types of fragmentation is also confirmed.
The capillary fragmentation of the jet occurs if the geometric parameter $L_j/D_j$ is above the critical value of $2$, and so independently of $We_d$. This finding is very robust, since it is the same for all studied cases (miscible, immiscible, totally and partially wetting jet liquid). Indeed, the only exception concerns the totally wetting drops, for which the fragmentation is observed at the same value, but only above a certain level of inertia.

Fragmentations attributed to an excess of inertia are also present. As with miscible liquids or immiscible totally wetting liquid jet,  the drops first fragment, marking the end of the  \textit{drops in jet} regime, followed for even greater inertia by the fragmentation of the jet, which leaves place to  \textit{mixed fragmentation}.

Yet, in contrast to the capillary limit for which a universal critical value of 2 is observed, for the inertial fragmentations shifts are observed in the associated values of $We_d$. Using silicone oil, the lower limit is found at $We_d\approx120$, significantly below  $We_d\approx150$, the limit observed for n-hexadecane,  see figures \ref{fig:G50_SOM5}(a) and  \ref{fig:G50_Hexa}(a). These observations call for a deeper analysis of the drop fragmentation process, which is taken in the next section.

\section{Drop and jet deformation, transition modeling}\label{mod}

%The inertial limits being non universally represented by a critical value of $We_d$ or a critical value of $We^*$, the modified Weber number proposed in \cite{PlanchetteBrenn2018}, we search for a new parameter. 
The inertial fragmentation of the drops and jet for partial and total wetting conditions are not described by the same critical value of $We_d$ or $We^*$, as introduced in \cite{PlanchetteBrenn2018}. This calls for the search of a new parameter able to universally describe this transition. To establish this parameter, it is important to keep in mind that inertio-capillary systems do not  always  evolve toward  the topology providing the minimum energy. Often, the final state of the system is actually selected by the kinetics of competing processes, such as pinch-off and recoil \citep{hoepffner2013, ilass_2019, hoath_2013}. Thus, we first focus on answering the following questions: how do the drops deform? Do they extend for a given $We_d$ to the same maxima with the same kinetics? Similar investigations are thus carried out for the jet, or more precisely for the liquid envelope of the system composed of the (transiently) merged drops and jet. After the maximal extensions have been reached, we focus on the recoil: by what is it governed?  Finally, we consider the drop fragmentation itself and search for the responsible mechanism.
Together with the results obtained for  the extension and the recoil phases, we propose a new parameter to model the drop fragmentation inside the jet. This parameter is then tested against the previously reported  data plus  additional data obtained with a fluorinated oil (see appendix \ref{appA} for more details about the perfluorodecalin experiments and data).

\subsection{Drop extension}\label{mod_D}
Qualitatively, the evolution of the drop  can be described as follows: upon impact,  it deforms and takes the shape of a bent lamella surrounded by a rim. During this phase, part of the drop kinetic energy is converted into surface energy.  The resulting lamella which can be modelled by a very flat disk (see section \ref{metandmat}) grows in diameter until it reaches its maximum $D^{max}_{lam}$. From this point on, %the internal kinetic energy is negligible, 
the  interfacial tension dominates, and the drop recoils to minimize its interfacial area. This aspect is treated later.

\begin{figure}
\centering 
  \includegraphics[width=\textwidth]{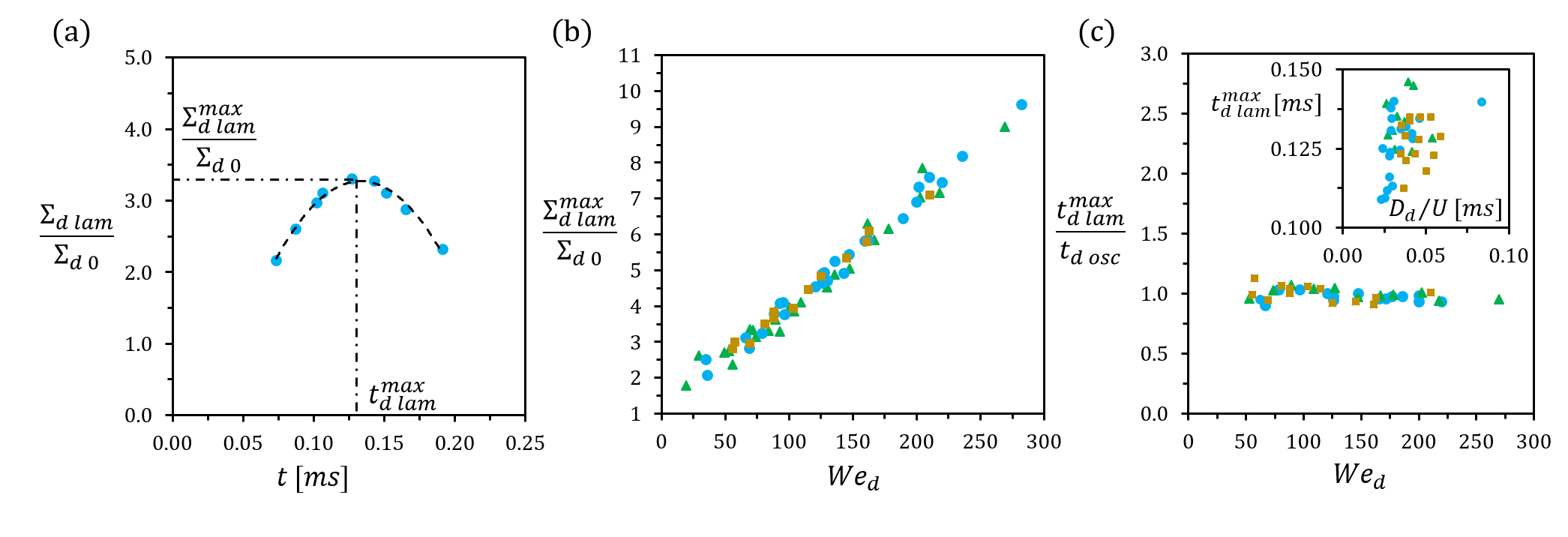}
  \caption{(a) Temporal evolution (blue circles: data points; dashed line: parabolic fit) of $\Sigma_{d\,lam}/ \Sigma_{d\,0}$ for G5 drops  and a n-hexadecane jet with  $We_d = 78$. The coefficient of determination $R^2$ is 0.9353. (b)  $\Sigma^{max}_{d\,lam} / \Sigma_{d\,0} $ as a function of $We_d$ for G5 drops with  a jet of: green triangles: SO 5; blue circles: n-hexadecane; and orange squares: perfluorodecalin. (c)  $t^{max}_{d\,lam}/t_{d\,osc}$ as a function of  $We_d$, inset $t^{max}_{d\,lam}$ as a function of $D/U$, same symbols as for (b).}  
\label{fig:sigma_dmax}
\end{figure}

To better understand how the drop deforms, we extract  for each collision the lamella surface, $\Sigma_{d\,lam}$, as a function of  the elapsed time $t$, starting from the instant of contact. The temporal evolution of $\Sigma_{d\,lam}$ is then fitted by a parabola, and the coordinates of the maximum $\Sigma_{d\,lam}^{max}$ and  $t_{d\,lam}^{max}$, the time period required for the drop to reach its maximal extension, are obtained, see figure \ref{fig:sigma_dmax}(a). We thus compute the maximum surface of the lamella $\Sigma_{d\,lam}^{max}$, which we normalize by the initial drop surface ${\Sigma}_{d\,0} = {D_d}^2 \pi$. Repeating this procedure for many collisions between drops of G5 and jets of silicone oil, n-hexadecane and perfluorodecalin, we obtain the results presented in figure \ref{fig:sigma_dmax}(b) showing $\Sigma_{d\,lam}^{max} / \Sigma_{d\,0} $ as a function of $We_d$. It is remarkable to observe that the evolution of $\Sigma_{d\,lam}^{max}/\Sigma_{d\,0}$ is independent from the liquid employed for the jet even though the drop gets encapsulated before reaching its maximal extension. The interfacial tension $\sigma_{d/j}$ plays surprisingly no role at this stage. Further, using $We_d$, all points collapse on the same straight line which is well fitted by $\Sigma_{d\,lam}° {max}/\Sigma_{d\,0}=  \alpha_{\Sigma_d} We_d + \beta_{\Sigma_d}$, in agreement with findings obtained experimentally  for drop-drop collisions  with one liquid  \citep{Willis-Orme_2003, Jiang-Umemura-Law_1992} or two immiscible ones \citep{PlanchetteBrenn2017} and  numerically for drop impact onto solid surfaces \citep{Wildeman2016}.
Here $ \alpha_{\Sigma_d}= 0.03$ and $\beta_{\Sigma_d}=1$, predicting, as expected, no deformation for a quasi-static approach of the jet by the drop.
In figure \ref{fig:sigma_dmax}(c), $t_{d\,lam}^{max}$ is plotted as a function of $t_{d\,osc}=\sqrt{\rho_d {D_d}^3/ \sigma_d}$, the capillary time. For the entire range of $We_d$, $t_{d\,lam}^{max}$  is proportional to $t_{d\,osc}$ and not to the inertial time scale $D_d/U$ (see inset). The coefficient of proportionality $c$ is equal to 0.33 for silicone oil and n-hexadecane, while a slightly smaller  value of 0.3 is found for perfluorodecalin. Similar findings were reported for drop-drop collisions with miscible and immiscible liquids \citep{PlanchetteBrenn2017} as well as for drop impacts onto solid surfaces \citep{Richard2002}. This purely capillary scaling has motivated modeling approaches where the drops are treated as liquid springs \citep{Okumura2003, PlanchetteBrenn2017} and inspired strategies to control the contact time between  drops and the solid surface on which they bounce \citep{Bird_2013}. 

Remarkably, these typical findings describing the drop extension are recovered in the present study, where drops impact onto a continuous immiscible jet.  Thus, the drop extension alone cannot explain the differences of critical values of $We_d$ associated to the  inertial fragmentation of the drops in the jet. Note that, for perfluorodecalin, its fragmentation is observed for $We_d \approx 95$, significantly below the values found for silicone oil and n-hexadecane ($120$ and $150$, respectively).
To go further, we now investigate the deformation of the jet interface,   i.e. of the liquid envelope.
\FloatBarrier

\subsection{Jet or envelope extension}\label{mod_d}
Detailed information about the estimation of the jet surface was given in section \ref{method}. After all relevant geometric parameters, see figure \ref{fig:dmax}(c) and (d), are extracted from the recorded pictures, the surface area is calculated according to eq. (\ref{surface_jmax}). For each collision, the calculated jet surface area $\Sigma_{j}(t)$ which corresponds to the area of jet  portion of length $L_j$,  is plotted as a function of time. After normalization of  $\Sigma_{j}(t)$  by its initial value $\Sigma_{j\,0}=\pi D_j L_j$, the data points are fitted by a third order polynomial function, similar as for the drops. For details, see appendix \ref{appC}. This type of function provides  good fits for all liquids within a defined time period around the maximum extension of the jet ($t_{j\,max}\pm{0.2}$ ms). We thus extract for each curve the value  of the local maximum, which provides $\Sigma_{j\,max}/ \Sigma_{j\,0}$. The time instant $t_{j\,max}$ is obtained by fitting the evolution of $D_{j\,ext (t)}/D_j$, which follows the same temporal evolution as $\Sigma_{j\,(t)}/ \Sigma_{j\,0}$, but subject to less measurement noise. See appendix \ref{appC} for a direct comparison.  Having done this fitting for many collisions involving different jet liquids, we test our results against appropriate scalings.

\begin{figure}
\centering
  \includegraphics[width=0.5\textwidth]{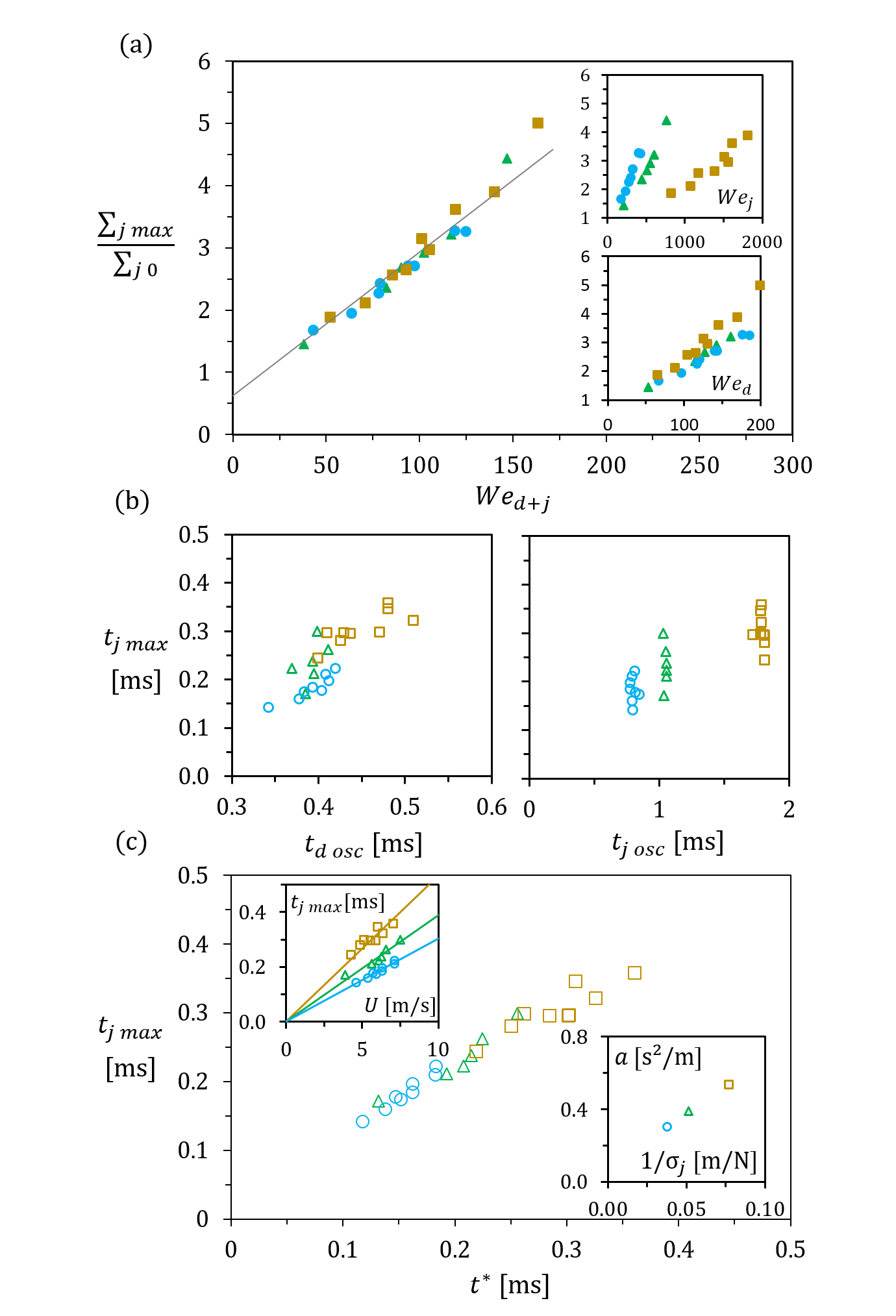}
  \caption{(a)  $\Sigma_{j\,max}/ \Sigma_{j{\,}0}$ (full symbols) as a function of $We_{d+j}$.  The grey line corresponds eq. (\ref{eq:jet_max}). The insets represent  $\Sigma_{j\,max}/ \Sigma_{j{\,}0}$ as a function of $We_j$ (top) and $We_d$ (bottom). (b)   $t_{j\, max}$ (empty symbols) plotted against the oscillating time period of the drop $t_{d\,osc}$ (left) and of the jet $t_{j\,osc}$ (right). (c) $t_{j\, max}$ versus $t^*= AU/\sigma_j$. The top inset evidences the linear variation of  $t_{j\, max}$ with $U$, the bottom inset shows the variations of $a$, the slope obtained for each jet liquid, as a function of $1/\sigma_j$. For all graphs: drops of G5 with a jet of  SO 5: green triangles; n-hexadecane: blue circles; and perfluorodecalin: orange squares.}
\label{fig:sigma_jmax}
\end{figure}

As expected, scalings based on $We_d=\rho_d D_d U^2/\sigma_d$ or $We_j=\rho_j D_j U^2/\sigma_j$ fail to bring all $\Sigma_{j\,max}/ \Sigma_{j{\,}0}$ measurements along a single line, shown in the insets of figure \ref{fig:sigma_jmax}(a). Instead, we observe a linear variation of $\Sigma_{j\,max}/ \Sigma_{j{\,}0}$ with either $We_d$ or $We_j$ for each liquid pair; the slope and the constant varying between silicone oil, n-hexadecane and perfluorodecalin. The surface scaling with $U^2$ points to a typical capillary-inertial dominated process. Yet, the slope and constant variations seem to indicate that at least one term - most likely the capillary one but possibly the inertial one - is wrongly evaluated by $We_d$ and $We_j$.
A careful analysis of the collision calls for a modified Weber number based on the ratio of the kinetic energy scaling as $\rho_d {D_d}^3 U^2$
and the surface energies opposing the deformation. These energies are composed of both the drop and jet contributions, namely $\sigma_d {D_d}^2 + \sigma_j S_j$. Here $S_j$ is the typical jet surface which is deformed by the collision. We evaluate this surface by $S_j=D_j D_d$. Here, $D_d$ is used to estimate the  typical length of the jet section impacted by the drop. Note that $L_j$ is too large to correctly represent this length ($L_j>D_j \approx 1.5 D_d)$. 
Thus we obtain for the modified Weber number:
\begin{equation}
We_{d+j}= \frac{\rho_d D_d U^2}{\sigma_d +\sigma_j D_j/D_d}
\end{equation}
The experimental data  $\Sigma_{j\,max}/ \Sigma_{j{\,}0}$ are plotted against $We_{d+j}$ in figure \ref{fig:sigma_jmax}(a). Independently from the jet liquid used, all points align along the same curve that can be well approximated by:
\begin{equation}
\Sigma_{j\,max}/ \Sigma_{j{\,}0} = \alpha_{d+j} We_{d+j }+ \beta_{d+j}
\label{eq:jet_max}
\end{equation}
where $\alpha_{d+j}=0.023 $ and $\beta_{d+j}=0.61$ (grey line).
We notice that $\beta \neq 1$, evidencing the validity limit of this scaling for very small inertia. Similar deviations were already observed for binary immiscible drop collisions and for ternary ones \citep{Planchette2012, PlanchetteBrenn2017}. While, to our knowledge,  they were not discussed in the literature, they could originate from the change of capillary energy due to the coalescence or encapsulation of the drops.
Similarly, for large kinetic energies, small deviations are observed which could be due to a transition toward a different regime of deformation, in which thinner and wilder structures may arise. Indeed, where such lack of linearity was reported for viscous binary drop collisions \citep{Willis-Orme_2003}, no explanation was provided. Finally, one cannot exclude  the deformation to be purely inertial. Expressing, instead of energy balance, the momentum conservation, an alternative scaling can be obtained which is also found to be very satisfying, see appendix \ref{appB}. Scalings derived from momentum conservation were successfully used in drop collisions studies, see for example \cite{Jiang-Umemura-Law_1992} and the  stretching separation limit established for binary drop collision. At this stage of our study, we do not have any strong argument to surely determine the most appropriate analysis.

%%%%%%%%%%here reformulate maybe leave scaling with We_d 

The time scale $t_{j\, max}$, at which the envelope reaches its maximum extension,  assimilated to the one for which $D_{j\,max}$ is maximum, is more difficult to apprehend. Looking at $t_{j\, max}$, we notice that neither  $t_{d\,osc}=\sqrt{\rho_d {D_d}^3/\sigma_d}$ nor $t_{j\,osc}=\sqrt{\rho_j {D_j}^3 /\sigma_j} $  represent well  the experimental data, see figure \ref{fig:sigma_jmax}(b). A modeling attempt is made by considering that the jet deforms as long as the kinetic pressure $p_k \propto U^2$  working on a typical section $\Sigma$, considered as fixed for constant $D_d$ and $D_j$, is sufficient. Balancing $p_k \Sigma$ with $\sigma_j (U t_{j\, max})$, the capillary force corresponding to the jet deformation, provides
\begin{equation}
 t_{j\,max}= p_k \Sigma / \sigma_j U \propto U/\sigma_j
 \label{eq:tjmax} 
\end{equation}
%. Here $V$ is the deformation velocity estimated by $U$. this provides: work to dthe work of the capillary forces corresponding to a given deformation corresponding to a certain kinetic energy $E_{k,j}$. We  write  $W_{\sigma j}= t_{j, max} V \sigma_j \delta_j$ with   $V$ the typical deformation  velocity and $\delta_j $  the typical length scale on which the capillary forces act. We further identify $V$ to $U$ %since 
%$(D_{d, max}-D_d)/t_{d,max} \propto (\sqrt{\alpha_{\Sigma_d}We_d+1}-1)D_d \sqrt{\sigma_d/\rho_d D_d^3} \approx  \sqrt{  \alpha_{\Sigma_d} We_d \sigma_d /(\rho_d D_d)} \propto U$.
%and consider that for a  given deformation $\delta_j$ is fixed (in this study the diameters of the drops and jet constant). By balancing this work with $E_{k,j}$, we obtain: 
%\begin{equation}
 %  t_{j, max} \propto \frac{U}{\sigma_j}  \bigg(  \frac{E_{k,j}}{\delta_j}  \bigg)=A \frac{U}{\sigma_j} 
%\label{eq:tjmax} 
%\end{equation}

%Since in this study the diameters of the drops and jet are kept constant ($D_d \approx 200 \mu m$; $D-j\approx 300 \mu m$), we simply this expression to obtain:    $t_{j, max}^{theo} =A \frac{U}{\sigma-j} $ 
We thus define $t^*= AU/\sigma_j$, where $A$ is a constant with the dimension kg/m, and plot the experimental data $t_{j\, max}$ against  $t^*$ with $A= 1500$ kg/m in figure \ref{fig:sigma_jmax}(c). The inset representing $t_{j\, max}$ as a function of $U$ demonstrates the linear variation (slope $a$ for each liquid). The second inset shows that the  relation between $a$ and $1/\sigma_j$ is according to the prediction of eq. (\ref{eq:tjmax}). Thus, despite the experimental uncertainty on $t_{j\, max}$, the agreement with eq. (\ref{eq:tjmax}) is excellent and only some slight deviations can be observed for collisions with fluorinated oil and large velocities. In the latter case, other effects could take place that we have neglected.

It is important to note that $t_{j \, max}$ is in all cases significantly larger than $t_{d\,lam}^{max}$. Indeed, for all liquids, we have reported $t_{d\,lam}^{max} \approx 0.3\, t_{d\,osc}$, (figure \ref{fig:sigma_dmax}(c)), while $t_{j\, max}>0.3\, t_{d\,osc} $ as evidenced in figure \ref{fig:sigma_jmax}(b)  for all investigated collisions. We also notice that the delay between  $t_{j\, max}$ and $ t_{d\, max} $  varies between the three jet liquids, in agreement with the finding of eq. (\ref{eq:tjmax}).  For a given impact velocity, this delay increases from n-hexadecane, to silicone oil and to perfluorodecalin. Further, for a given liquid, the delay increases with $U$. 

Since in all investigated collisions, the drop has already started to recoil while the jet keeps extending, it is legitimate to compare the previous time scales,  $t_{j\, max}$ and $t_{d\,lam}^{max}$,  to the instant  $t_{frag}$ at which the drop fragments. Is the drop fragmentation driven by the extension or by the recoil of the drop or jet?

\subsection{Drop pinch-off, on the importance of recoil}\label{mod_zeta}

To investigate the drop fragmentation, its mechanisms and kinetics, pictures recorded with camera 1 providing the so-called orthogonal view are used. Thanks to the dye present in the drop liquid only, we can follow the evolution of the drop and observe its potential fragmentation.

\begin{figure}
\centering
  \includegraphics[width=\textwidth]{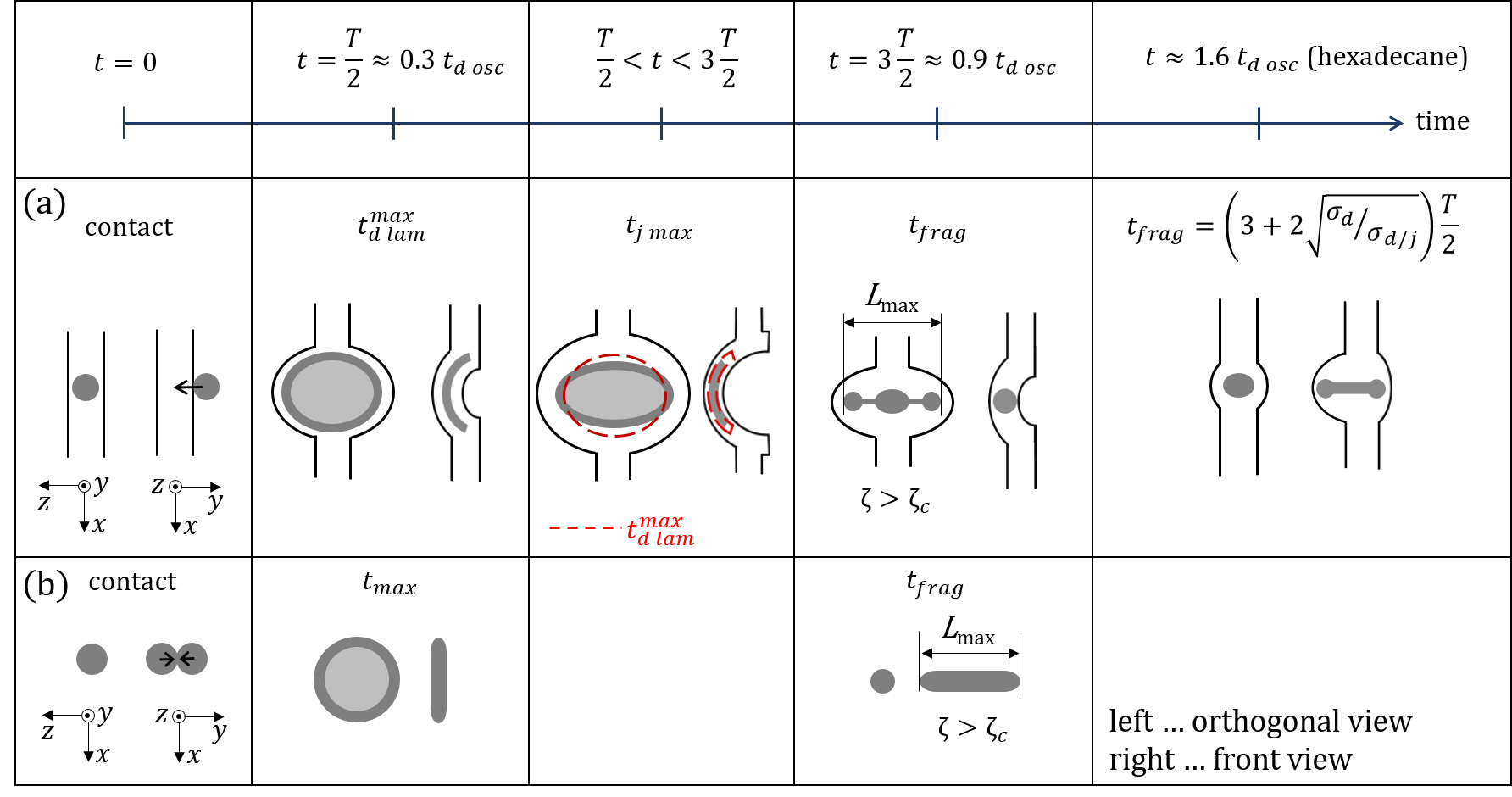}
  \caption{(a)  Schematic representation of a drop-jet collision at $t^{max}_{d\,lam}$, the instant of drop maximal  extension (lamella);  $t_{j\,max}$, the instant of jet reaches  maximal extension (the red dashed lines mark the  drop maximal extension) and  $t_{frag}$, the instant of drop fragmentation (if $\zeta=L_{max}/D_d > \zeta_c$, introduced thereafter). (b)   Schematic representation of a drop-drop collision at $t_{max}$, the instant of drop maximal  extension and $t_{frag} $ the instant of  fragmentation (when $\zeta > \zeta_c$). The sketches show for each instant a front (right) and orthogonal (left) view of the system.}
\label{fig:sketch_collision}
\end{figure}

To understand the evolution of the drop recoil within the jet, it may be useful to recall results obtained with drop impacts on solid surfaces or drop-drop collisions. In the latter situations, the drop  recoil is isotropically  directed toward the center of the lamella, forming a cigar whose axis is aligned with the one of the lamella, see  figure  \ref{fig:sketch_collision}(b). This cigar may fragment following a pseudo Rayleigh criterion and giving raise to two  drops of equal size.  For drop-jet collisions, the constraint exerted by the jet interface onto the encapsulated drop dramatically modifies the recoil geometry as sketched in figure  \ref{fig:sketch_collision}(a). While the drop keeps forming a (bent) lamella which relaxes to produce  an elongated drop, the axis of the elongated drop (along its length $L_{max}$) is found to be perpendicular to the collision plane, i.e. perpendicular to the orientation it would take in the absence of the jet. When the drop fragments, it occurs on both extremities of this elongated drop according to a pinch-off process resulting in a main drop and two smaller satellite drops.

\begin{figure}
\centering
  \includegraphics[width=\textwidth]{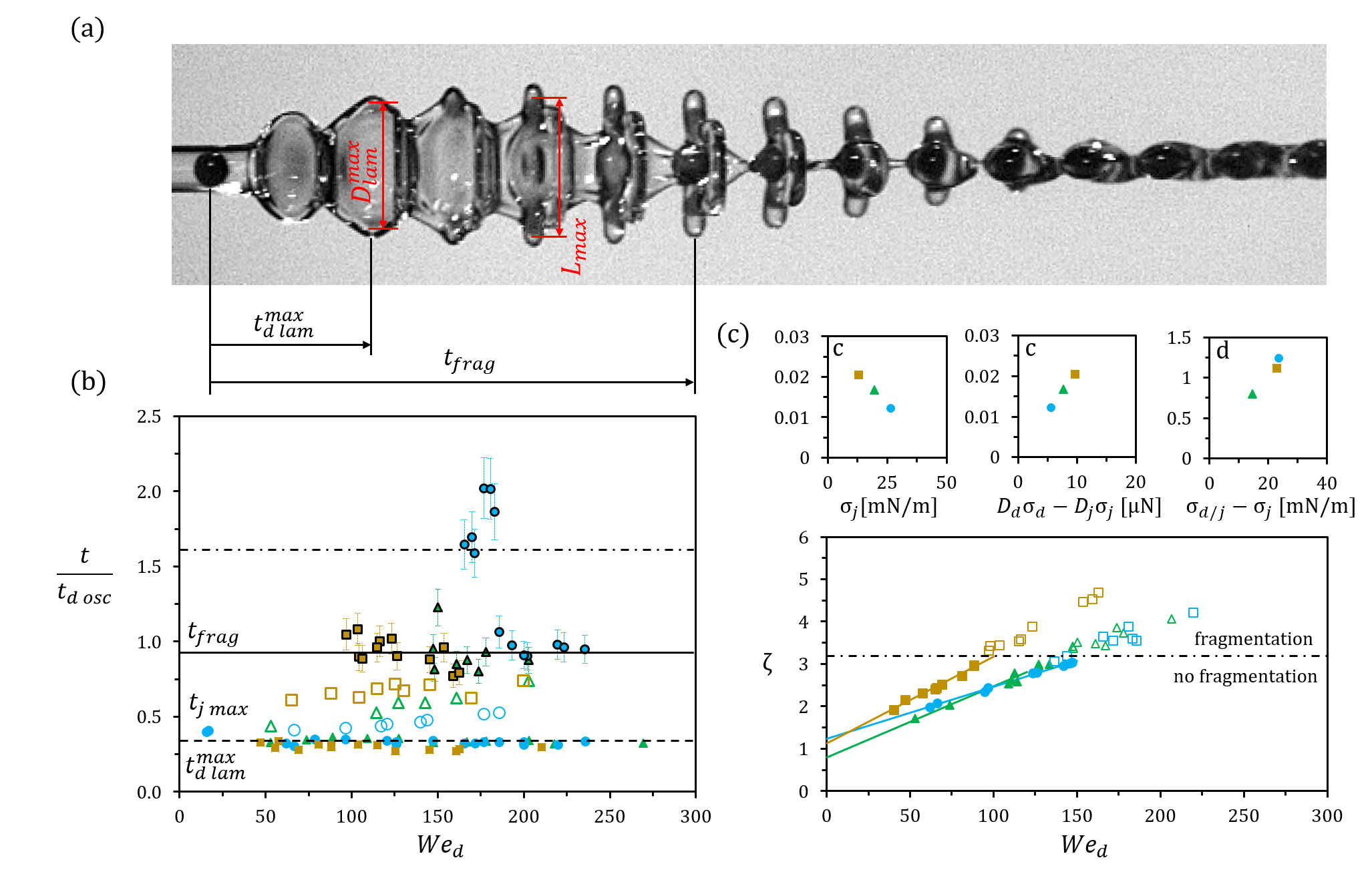}
  \caption{(a) Collision between G5 drops and SO 5 jet at $We_d=178$ and $L_j/D_j=1.33$ recorded with the orthogonal camera. Schematic illustration of $L_{max}$, $D^{max}_{lam}$, $t^{max}_{d\,lam}$ and $t_{frag}$. (b) Time scales of the collision with $t^{max}_{d\,lam}$: full color symbols, $t_{j\,max}$: empty color symbols and $t_{frag}$): black-color symbols as a function of $We_d$. All time scales are normalized by  $t_{d\,osc}$ and the jet liquids are: blue: n-hexadecane; green: silicone oil; and orange: perfluorodecalin.   (c) The normalized length of the elongated drop $L_{max}/D_d =\zeta$ as a function of $We_d$ using full symbols for \textit{drops in jet} and open ones for \textit{fragmented drops in jet} or \textit{mixed fragmentation}. The small graphs illustrate the slope $c$ of the curves subjected to $\sigma_j$ and $D_d\sigma_d-D_j\sigma_j$ as well as the constant $d$ as a function of $\sigma_{d/j}-\sigma_j$.}
\label{fig:drop_frag}
\end{figure}

To quantify this process, especially its kinetics, we report in figure \ref{fig:drop_frag}(b)  the instant of fragmentation normalized by the drop oscillation time, $t_{frag}/t_{d\,osc}$, as a function of $We_d$. For comparison we also plot $t_{d\,lam}^{max}/t_{d\,osc}$ and $t_{j\,max}/t_{d\,osc}$. To give insight on the fragmentation threshold, and thus on its mechanism, we also report the normalized length of the elongated drop $L_{max}/D_d =\zeta$ as a function of $We_d$, illustrated in figure \ref{fig:drop_frag}(c).
Here as well, the data are presented for three jet liquids (silicone oil, n-hexadecane and perfluorodecalin), the drop liquid remaining unchanged (G5).
We observe that, independently from the jet liquid, the drop fragmentation takes place after the maximal extension of both the drop  and  the jet, $t_{d\,lam}^{max}$ and $t_{j\, max}$, approximately after $3\,t_{d\,lam}^{max}$ except for a few points obtained with n-hexadecane and for which the fragmentation occurs even later, around $6\,t_{d\,lam}^{max}$ (see fifth column in figure \ref{fig:sketch_collision}).  %error bars ($\pm{10}$ \%) are added to $t_{frag}$ in the graph \ref{fig:drop_frag}(b) 
Note that, since the exact instant of fragmentation can neither be exactly  determined from the pictures nor deduced from a fit (in contrast to $t_{d\,lam}^{max}$ and $t_{j\, max}$), error bars have been added to $t_{frag}$, figure \ref{fig:drop_frag}(b). They correspond to the last instant  at which the drop is definitely seen as not fragmented and to the first instant where the drop can confidently be  observed as fragmented.
\,\,Yet, it is worth commenting the finding that $t_{frag} \approx 3\,t_{d\,lam}^{max} \approx 0.9\, t_{d\,osc}$. The late fragmentation cases observed with n-hexadecane will be discussed at the end of this paper. First, beside the presence of different encapsulating liquids, and more importantly beside the changes adopted by the recoiling drop, this result is similar to the one obtained for drop-drop collisions using a single liquid \citep{Willis-Orme_2003}, see also figure \ref{fig:sketch_collision}(b). This seems to indicate  that the kinetics of the drop recoil remains dominated by the typical oscillation time $t_{d\,osc}$ of an isolated drop  and that the pinch-off is driven by an excessive extension of the recoiling drop, as for drop-drop collisions with two and three drops, using one or two liquids \citep{Willis-Orme_2003, Planchette2012, Hinterbichler-Planchette-Brenn_2015, PlanchetteBrenn2017}.
This hypothesis can be further tested by looking at figure \ref{fig:drop_frag}(c), where $\zeta$ is plotted as a function of $We_d$ using full symbols for \textit{drops in jet} outcomes and open ones for \textit{fragmented drops in jet} or \textit{mixed fragmentation}. While for each liquid pair, the data aligned along different lines, the transition between full and empty symbols is found at the same critical value of $\zeta$ ($\zeta_c \approx 3$) for all liquids. Thus, attributing the pinch-off of the encapsulated drop to an excessive elongation appears relevant. The critical value is close to the ratio of disturbance wavelength to filament diameter separating the unstable from the stable ranges of wavelengths, as predicted by Rayleigh\textquotesingle s liquid jet stability analysis. This break-up mechanism and critical value were mentioned for other liquid systems deformed by collisions  \citep{PlanchetteBrenn2017}. However, in the present study, the shape of the elongated drop strongly differs from a cylinder evoking a different break-up mechanism.
A closer inspection of the drop at fragmentation shows  dumbbell-shaped drop endings. As already mentioned, these bulbs may  then pinch off, leaving a main drop and two smaller ones, strongly recalling  the end-pinching mechanism introduced by \cite{Stone1986} and \cite{Stone1989}. We note that the critical value of $3$ is slightly lower than 4, the value found in the absence of flow for the end-pinching mechanism \citep{Stone1986}. Further experiments (data not shown) indicate that increasing or decreasing the viscosity ratio between the drop and jet phases leads to larger values of $\zeta_c$ as expected for the mentioned pinch-off mechanism. Thus, the discrepancy is most probably caused by complex flows and complex drop shape. While flows within the encapsulated phase may delay the pinch-off \citep{hoepffner2013}, it was reported that flows in the encapsulating phase lead to break-up for  smaller drop extension   \citep{stone_leal_1989}. In the current process, the jet envelop is  subjected to capillary recoil which generates non negligible  flows. Combined with the drop-jet geometry, this recoil further influences  the shape of the encapsulated drop which differs from the cylinder considered in \cite{Stone1986, Stone1989}. Indeed, the lamella is drained into two opposite bulbs separated by a central part. The corresponding local curvatures  are expected to generate additional capillary instabilities and could therefore cause the pinch-off to occur for lower values of $\zeta_c$.

It is also remarkable to observe that, for each liquid,  $\zeta$ seems to vary linearly with $We_d$, the slope and constant varying from liquid to liquid. This indicates that the elongation of the drop during its recoil phase has in all cases an inertial origin. Finally, it can also be noticed that the values measured above $\zeta_c$ are more dispersed than those found below. This can be understood as a consequence of the drop pinch-off that may slightly affect the measurement of $L_{max}$, adding noise to the data.

At this stage of our analysis, one question remains: what fixes the slope and constant of the linear variation of $\zeta$ with $We_d$? Or in different terms, how to predict the evolution of $\zeta$ to deduce the corresponding value of $We_d$ marking the first inertial fragmentation?
 
\subsection{Consequences on the inertial fragmentation limit}\label{mod_map}

We extract from figure \ref{fig:drop_frag}(c) the slope $c$ and the constant $d$ for each curve, so that for a given liquid pair the evolution of $\zeta$ with $We_d$ is well reproduced by $\zeta= c We_d + d$. We attribute the term in $c We_d$ to the relative deformation and recoil of the drop and jet and thus expect both $\sigma_d$ and $\sigma_j$ to influence the value of $c$. This interpretation is tested in figure \ref{fig:drop_frag}(c).%, where $c$ is plotted as a function of $\sigma_j$ and of $\sigma_d D_d -\sigma_j D_j$. 
The larger $\sigma_j$, the smaller $c$, indicating that the lateral extension of the drop is limited by large values of $\sigma_j$. Excellent agreement is found by comparing $c$ to $\sigma_d D_d -\sigma_j D_j$, where $D_d=200$ \textmu m and $D_j= 300$ \textmu m. Beside the tentative character of our analysis, the expression $\sigma_d D_d -\sigma_j D_j$ seems to confirm the competitive character of the drop and jet deformation and recoil.  
The constant term $d$ is independent of the deformation magnitude (independent of $We_d$) and is thus attributed to the relative deformability of the jet and  the drop caused by the relative Laplace pressures. Indeed, whatever the deformation amplitudes are, the interface of  the encapsulated drop and the one of the jet envelope follow each other, so that, at first order, the curvatures coincide. Thus, the difference between the Laplace pressure jumps at the jet and at the encapsulated drop is expected to be proportional to the difference of interfacial tensions $\sigma_{d/j}-\sigma_{j}$. This is in very good agreement with the experimental measurements, as shown by the graph representing $d$ as a function of $\sigma_{d/j}-\sigma_{j}$. Thus we obtain: 
\begin{equation}
    \zeta^{theo} = 2.9 \cdot 10^{-2} (1 -\sigma_j D_j /\sigma_d D_d) We_d + 3.5 (\sigma_{d/j}-\sigma_{j})/ \sigma_d
    \label{eq:ccl}
\end{equation}
\begin{figure}
\centering
  \includegraphics[width=\textwidth]{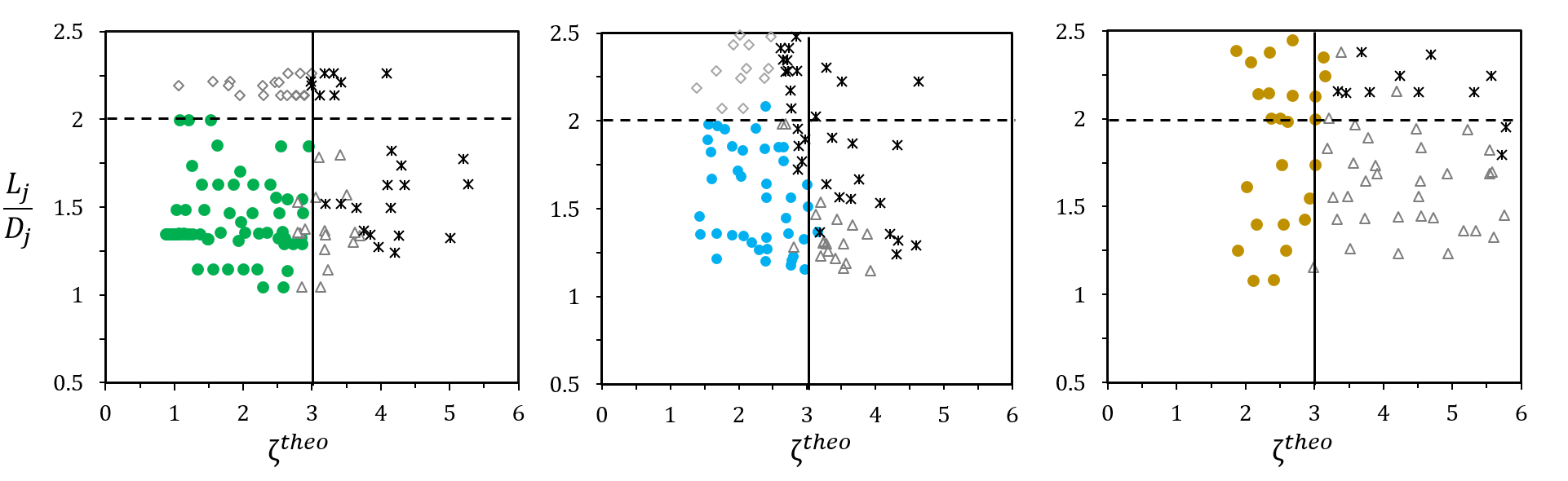}
  \caption{Regime maps with $L_j/D_j$ and $\zeta^{theo}$ as scaling parameter. G5 is used as drop liquid and as jet liquid from left to right: silicone oil, n-hexadecane and perfluorodecalin. Full circles: \textit{drops in jet}, empty diamonds: \textit{encapsulated drops}, empty traingles: \textit{fragmented drops in jet} and black crosses: \textit{mixed fragmentation}. }
\label{fig:zeta_theor}
\end{figure}

Here, the surface and interfacial tensions have been normalized by the one of water, and the dimensions by  the drop diameter $D_d$. The numerical factors $2.9 \cdot 10^{-2}$ and $3.5$ are found empirically from the graphs of figure \ref{fig:drop_frag}(c) using all points below the drop fragmentation threshold, i.e for $40<We_d<160$. Their validity  cannot be determined with currently available data. Further experiments involving other liquid pairs should be performed to address this point. These models (eq. \ref{eq:ccl}) are used in figure \ref{fig:zeta_theor}, where ($L_j/D_j$; $\zeta^{theo}$) regime maps are plotted for, from left to right: silicone oil, n-hexadecane and perfluorodecalin. For all jet liquids, the first inertial fragmentation (vertical continuous line) is  found for a critical value of $\zeta^{theo}=3$, in agreement with our analysis. The other capillary limit remains unchanged and is found for $L_j/D_j \approx 2$ (dashed lines) for all jet liquids. 

Thus, to account for different surface and interfacial tensions, we  propose to replace the inertial-capillary parameter $We^*$ introduced in \cite{PlanchetteBrenn2018} or $We_d$ by  $\zeta^{theo}$. In contrast to $We^*$ or $We_d$, it enables to account for the competition between the  drop and jet inertial extension via  $(1-\sigma_j D_j/\sigma_d D_d) We_d$. Further, the term $(\sigma_{d/j}-\sigma_j)/\sigma_d$  renders the relative deformability of the jet envelope and the encapsulated drop during recoil. This parameter has been built based on a detailed analysis of drop-jet collisions produced by strongly varying the interfacial tension as well as the jet surface tension and density. By testing this parameter against the three immiscible jet liquids of this study (silicone oil, n-hexadecane and perfluorodecalin) we demonstrate its robustness for predicting the first inertial frgamentation. We recall that, in this study, liquids were chosen to have similar viscosity and viscosity ratio, which is the reason why they do not appear in $\zeta^{theo}$. Indeed, we expect viscous losses to strongly modify the amplitude of the drop and jet extensions \citep{PlanchetteBrenn2018},  the critical aspect ratio of pinch-off \citep{Stone1986}, and more generally the kinetics of  deformation. Additionally, strong variations in the size ratio between the drops and jet may significantly modify the qualitative evolution of the drops and jet, which constitutes another limit to the present analysis.  These aspects go beyond the scope of this study and will be presented in a forthcoming article. 

Finally, let us come back on the few late fragmentation cases observed for n-hexadecane, for a moderate $We_d$, i.e. between the point where the drops start to fragment and $We_d\approx185$. Note that, above this value of $We_d$,  the kinetics observed for other liquids is recovered, and fragmentation occurs around $3\, t_{d\,lam}^{max} \approx 0.9\,t_{d\,osc}$. For lower $We_d$, the drop fragmentation does not happen around $0.9\, t_{d\,osc}$,  but around $1.6\,t_{d\,osc}$ and later. By carefully looking at the pictures  taken by both the front and side cameras, we notice that the first elongation of the drop still occurs around $3\,t_{d\,lam}^{max} \approx 0.9\, t_{d\,osc}$, but without leading to immediate pinch-off. Instead, the drop further recoils and extends in the orthogonal direction. The fragmentation takes place  during this second orthogonal elongation. The absence of fragmentation during the first elongation could be first interpreted as the result of internal flows \citep{Stone1986}. Yet, this does not happen for other liquids, while the internal flows are expected to be similar (for similar deformation at least). Thus, another reason could be the fast recoil kinetics of the liquid system. Indeed, for n-hexadecane, %the interfacial tension between the drop and jet liquids is approximately $\sigma_{dj}=50 mN/m$, much larger than the ones found for silicone oil ($\sigma_{dj}= 34 mN/m$) and perfluorodecalin ($\sigma_{dj}=36 mN/m$). Similarly, 
the jet surface tension $\sigma_{j}=26.5$ mN/m is larger than for silicone oil and perfluorodecalin ($19.5$ mN/m and $13$ mN/m, respectively). Higher surface tension  leads to a faster contraction of the jet envelopes. For moderate excess of kinetic energy (moderate $We_d$), and thus moderate elongation $L_{max}/D_d$, the jet recoil kinetics may be too fast and may prevent the encapsulated drop   pinching off. Instead, the drop fragmentation occurs during the second elongation, whose kinetics may be slowed down. Assuming that, after the first drop recoil, the drop behaves as a spring whose constant is fixed by its interfacial tension $\sigma_{d/j}$, we find that the oscillation period of the encapsulated drop  is modified by a factor $\sqrt{\sigma_d/\sigma_{d/j}}=1.16$, providing $t_{d\,osc}^*= 1.16\, t_{d\,osc}$. The fragmentation can therefore be expected for $t_{frag}^*=  0.3 (3\,t_{d\,osc} + 2\,t_{d\,osc}^*) \approx 1.6 \,t_{osc}$, see figure \ref{fig:sketch_collision}(a). This value is in good agreement with the experimental results (see figure \ref{fig:drop_frag}(b), dashed-doted line). This interpretation is also coherent with previous analysis and would explain why this phenomenon is observed only for n-hexadecane and moderate $We_d$. Note that we cannot exclude further causes to the kinetic changes. While viscous friction between the two phases may not be a good candidate (n-hexadecane is slightly less viscous than silicone oil and perfluorodecalin), a stronger Laplace pressure jump may change the internal flows, making the pinch-off possible only during the second elongation. 

\FloatBarrier                    

\section{Summary and conclusions}\label{conc}
Collisions between a drop stream and  a continuous liquid jet were experimentally investigated, especially focusing on the effects of liquids miscibility and wettability. For all liquids and liquid pairs, head-on collisions were considered between  droplets  of diameter $D_d=200\pm{20}$ \textmu m and a jet of diameter $D_j=290\pm{20}$ \textmu m. The normalized spatial frequency of the collisions  ${L_j}/{D_j}$ was varied between $1.0$ and $2.5$. The relative velocity between the drops and jet was found between 2 m/s and 10 m/s.
 
Various liquids were used to probe the effects of liquid wettability and miscibility.

First, using immiscible liquids with total wetting of the drop liquid by the jet liquid (drops  of an aqueous glycerol solution and jet of   silicone oil), various collision outcomes were observed and categorized in  four regimes, depending if the drops only, the jet only, both or none of them were fragmenting after the collisions. 
Two fragmentation mechanisms were identified: a   fragmentation of capillary origin responsible for the jet  break-up above a critical value of $L_j/D_j\approx 2$, and an inertial fragmentation causing first the break-up of the drops  above a critical value of $We_d\approx 120$.

This qualitative description was confronted to the observations made with other liquid pairs. Using miscible liquids (drops  of an aqueous glycerol solution and jet of an aqueous ethanol and glycerol solution), and beside the absence of interface between drops and jet, similar outcomes were observed. Indeed, by redefining the regimes based on the liquid spatial  distribution after the collisions (drop or jet liquid found in more or less entities than before the collision), a similar ($L_j/D_j$ ; $We_d$) regime map could be established. The capillary limit found at $L_j/D_j \approx 2$ is still visible and the inertial limit remains similar with  critical value of $We_d$ of 120.

In contrast to the surprisingly small differences caused by liquid miscibility, reversing the drop and jet liquids (drops   of  silicone oil and jet of an aqueous glycerol solution) caused significant changes in the collision outcomes. The  qualitative description previously used does not hold any more. Indeed,  within the studied  parameter ranges, it is impossible to encapsulate the drops within the jet. Instead the drops spread around the jet, forming a coaxial structure. Further,  the capillary limit previously observed for $L_j/D_j \approx 2$ partly disappears. It is recovered for large enough values of $We_d$ (500 for  $L_j/D_j > 2$ and 700 for $L_j/D_j < 2$), but with an extremly different  liquid distribution.  Thus, considerable care should be given to the relative value of the drop and jet surface tensions when aiming to use these collisions to produce capsules or fibers. 

Finally, various jet liquids are used whose surface tension is always smaller than the one of the drops (aqueous glycerol solution). Additionally to silicone oil, perfluorodecalin is used to provide total wetting, while n-hexadecane leads to partial wetting of the drops by the jet. 
Qualitatively, the regime maps obtained with these three liquid pairs are very similar, encouraging the use of such collisions to encapsulate drops in a jet. In all cases, we observe the four regimes found for silicone oil separated by the two fragmentation mechanisms of capillary and inertial origin. The capillary limit is very robust and always found for $L_j/D_j \approx 2$.
The inertial fragmentation limit,  in contrast, is found for various values of $We_d$ ranging from 90 for perfluorodecalin up to 150 for n-hexadecane. This important result,   defining the range of collision parameters enabling the production of encapsulated drops in a jet, leads us to a detailed analysis of the collision.

Our main findings show that the drop maximal extension (amplitude and kinetics) is governed by $We_d$ and $t_{d\,osc}$, the oscillation period of the drop, independently from the jet liquid used and similarly to the results obtained for  drop impacts onto solid surfaces and drop-drop collisions. The jet extension  also linearly increases with $We_d$ but is further affected by the jet surface tension $\sigma_j$ indicating a probable capillary-inertial origin. Further, and in contrast with what is found for the drop,  the instant of maximal extension increases with the inertia of the system. Finally, the drop fragmentation occurs after both the drop and the jet have reached their maximal extension. The shape of the recoiling drop is constrained by the interface of the jet - itself recoiling - which strongly modifies its shape by comparison to drop impact on solid surfaces or drop-drop collisions. Yet, for all jet liquids, the same critical value of $\zeta$, the normalized drop elongation, can be associated to its fragmentation. This fragmentation seems  to happen rather according to an end-pinching process than to some capillary instability. Proposing a semi-empirical law for the aspect ratio of the elongated drop, we unify all inertial fragmentation limits with $ \zeta^{theo}_{c} = 2.9 \cdot10^{-2} (1 -\sigma_j D_j /\sigma_d D_d) We_d + 3.5 (\sigma_{d/j}-\sigma_{j})/ \sigma_d =3$. This newly defined parameter will be useful for the exploitation of drop-jet collisions as an encapsulation method. %Further, we hope that  the detailed study of the drop and jet extension and kinetics will be inspiring for similar yet different drop collisions (on liquid bath, films, immiscible drops...)

\section*{Declaration of Interests }
None.

\section*{Acknowledgements}
DB, CP and GB gratefully acknowledge the financial support of the Austrian Science Fund (FWF) for the project number P 31064-N36. All authors gratefully acknowledge the financial support of the Deutsche Forschungsgemeinschaft (DFG) within the frame of the DROPIT (GRK2160/1) summer school 2018. We further want to thank Prof. Gescheidt-Demner (TU Graz, Institute of Physical and Theoretical Chemistry), Prof. Siebenhofer with technician T. Weiss (TU Graz, Institute of Chemical Engineering and Environmental Technology) and Dr. W.K. Hsiao (RCPE GmbH, Graz) for their technical support and advice regarding the surface and interfacial tension measurements.

\FloatBarrier
\appendix
\section{Jet liquid: perfluorodecalin}\label{appA}
\begin{figure}
\centering
  \includegraphics[width=\textwidth]{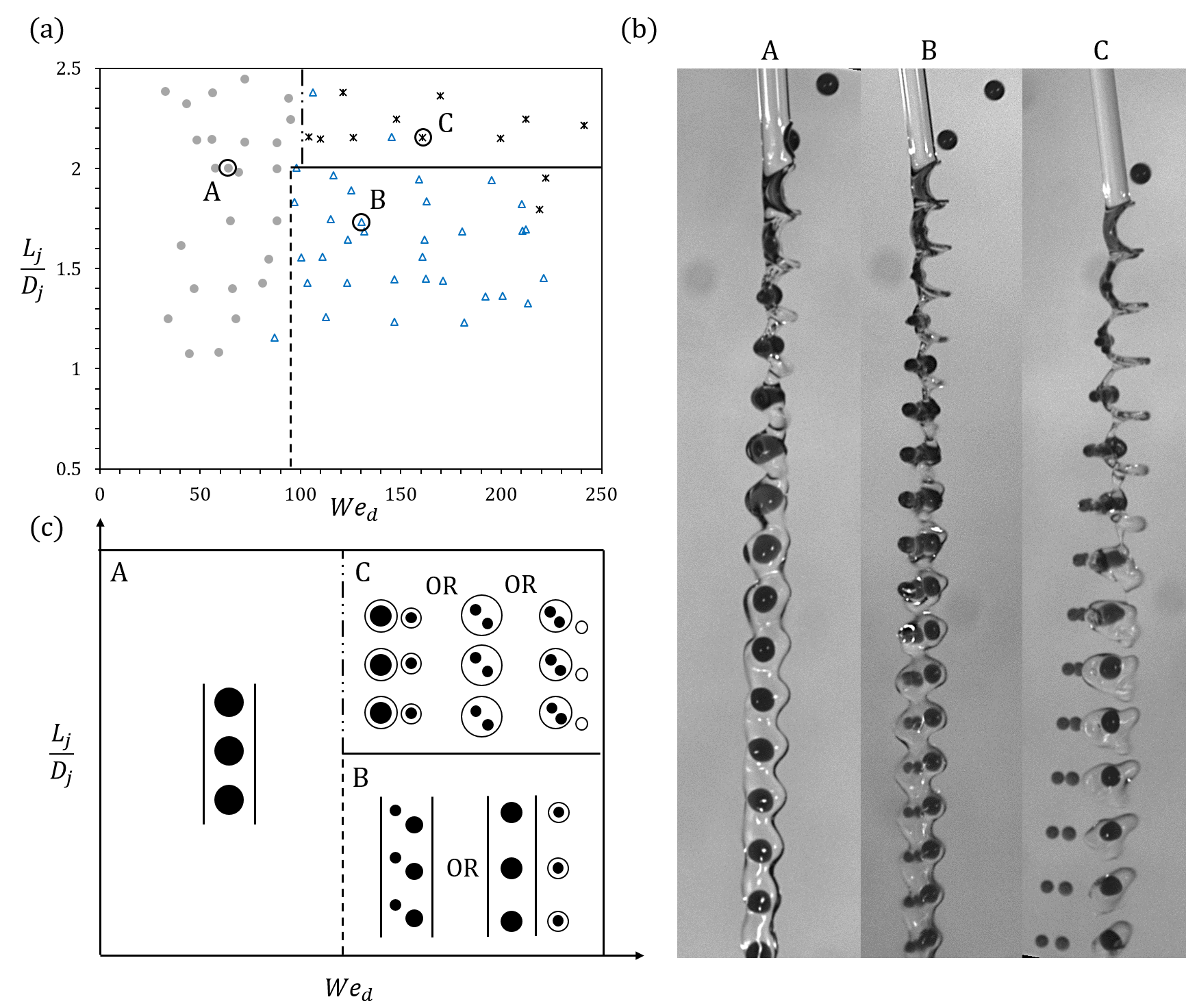}
  \caption{(a) Experimental regime map of  G5 drop  and PFD jet. (A) \textit{drops in jet} full gray circles,  (B) \textit{fragmented drops in jet} empty blue triangles and (C)  \textit{mixed fragmentation} black crosses. The lines are guides for the eye, the black circles correspond to the pictures of part (b). (b) Collisions pictures with (A) ${L_j}/{D_j}=2.00$, $We_d=63.75$; (B) ${L_j}/{D_j}=1.68$, $We_d=131.71$ and (C) ${L_j}/{D_j}=2.15$; $We_d=126.43$. (c) Schematic representation of the observed regimes. Black: drop liquid (G5), white: jet liquid (PFD).}
\label{fig:G50_PFD}
\end{figure}

Figure \ref{fig:G50_PFD} shows (a) the regime map, (b) illustrative  pictures and (c) sketches of the regimes obtained with perfluorodecalin (PFD) as the jet liquid and the glycerol solution G5 as the drop liquid. The drop is  coloured with blue dye, while the jet stays transparent. The two liquids are immiscible and PFD totally wets G5. The density of PFD is about twice the density of G5, as listed in table \ref{tab:fluidproperties}. The normalized period ${L_j}/{D_j}$ varies between $1.0$ and $2.5$ and the relative impact velocity $U$ ranges from 2 $m.s^{-1}$ to 10 $m.s^{-1}$. Within these parameter ranges, only three regimes can be observed, as shown in figure \ref{fig:G50_PFD}:
\begin{itemize}
\item$\,$\textit{drops in jet}: The drop and the jet remain stable after the collision, and the drop is totally encapsulated by the jet liquid. %A continuous, cylindrical jet with a regular embedded drop stream is the result, as shown in 
 Note that, in contrast with silicone oil (section \ref{sec:G5_SOM5}), no direct transition toward \textit{encapsulated drops} can be observed for small $We_d$ and $L_j/D_j>2$.  %  It is not possible to break the jet only due to a capillary mechanism at low kinetic energy of the drop. The critical value of 2 for the capillary fragmentation limit only holds for higher impact velocity of the drop. The jet is too massive in order to easily break the jet due to the induced disturbances with low kinetic energy of the drop.
The inertia dominated fragmentation limit is found again for smaller $We_d$ ($We_d\approx95$) than observed with silicone oil and n-hexadecane.    
 See figure \ref{fig:G50_PFD}(a-b-c)-A.
\item$\,$\textit{fragmented drops in jet}, identical to that observed with silicone oil and n-hexadecane, is observed. The drops fragment inside the jet which remains continuous. In some cases, and as already mentioned for the other liquids,  drop fragments may be expelled from the jet leading to a continuous jet with embedded drops accompanied by a regular stream of satellites drops. The regime starts at $We_d\approx95$ and can be observed for all values above this limit. The capillary limit found with other liquids at $L_j/D_j=2$ is recovered and marks the transition to \textit{mixed fragmentation}. See figure \ref{fig:G50_PFD}(a-b-c)-B.
\item$\,$Finally, the \textit{mixed fragmentation} regime, corresponding to both  drop and jet fragmentation, can be seen. See figure \ref{fig:G50_PFD}(a-b-c)-C.  
\end{itemize}    
\FloatBarrier

\section{Change of reference system and jet deformation}\label{appB}
As an alternative approach to the one presented in the main text, one can model the deformation of the jet based on momentum conservation. As classically done with perfectly inelastic collision, the momentum conservation can be expressed using the referential of the center of mass of the system "drop + jet".  The change of referential is presented below. 

Figure \ref{fig:change_refsystem} represents the kinetic parameters of the collision in the reference system $x$,$y$,$z$ placed on the jet, and in the reference system $x^*$,$y^*$,$z^*$ placed on the centre of mass (subscript: CM). %All needed parameters for the following calculation steps can be found in figure \ref{fig:change_refsystem}. 
The relative velocity $\vec{U}$ between drop and jet is independent from the chosen referential and given by:  
\begin{equation}
\vec{U}=\vec{u_d}-\vec{u_j}
\label{eq:Urel_vector}
\end{equation}
%Its norm is defined as: 
%\begin{equation}
%U=\sqrt{(u_d\,\cos{\alpha}-u_j)^2+(u_d\,\sin{\alpha})^2}
%\label{eq:Urel_abs}
%\end{equation}

Applying the momentum balance between the drop and the jet provides the center of mass velocity: %a quantitative description of the referential  $x^*$,$y^*$,$z^*$.
%\begin{equation}
%x^*{:}\qquad (m_d+m_j)u_{CM}=m_ju_j\cos{\gamma}+m_du_d\cos{(\alpha-\gamma)}
%\label{eq:momentum_xstar}
%\end{equation}

%\begin{equation}
%m_j \vec{u_j} + m_d \vec{u_d} = (m_j+m_d) \vec{u_{CM}}
%\label{eq:momentum}
%\end{equation}

%which provides

\begin{equation}
\vec{u_{CM}}=\frac{m_j }{m_j+m_d} \vec{u_j} + \frac{m_d}{m_j+m_d}  \vec{u_d} 
\label{eq:momentum_2}
\end{equation}

%\begin{equation}
%y^*{:}\qquad 0=m_ju_j\sin{\gamma}-m_du_d\sin{(\alpha-\gamma)}
%\label{eq:momentum_ystar}
%\end{equation}  
%Equation (\ref{eq:momentum_ystar}) can be rewritten as:
%\begin{equation}
%\tan\gamma=\frac{\sin\alpha}{({m_ju_j})/({m_du_d})+\cos\alpha}
%\label{eq:tan_gamma}
%\end{equation} 
With the help of equation (\ref{eq:momentum_2}),  the relative velocity between the drop and the centre of mass, $\vec{U}_d^*$, %and  the relative velocity between the jet and the centre of mass $\vec{U}_j^*$ are:
writes :
\begin{equation}
\vec{U}_d^*=\vec{u}_d-\vec{u}_{CM}=\frac{m_j}{m_d+m_j} (\vec{u_d}- \vec{u_j})= \frac{m_j}{m_d+m_j} \vec{U}
\label{eq:transf_coeff_Ud}
\end{equation}

Similarly, we obtain:
\begin{equation}
\vec{U}_j^*=\vec{u}_j-\vec{u}_{CM}=\frac{m_d}{m_d+m_j} \vec{U}
\label{eq:transf_coeff_Uj}
\end{equation}

%respectively. The norms of $\vec{U}_d^*$ and $\vec{U}_j^*$ can then be calculated:
%\begin{equation}
%U_d^*=\sqrt{(u_d\cos{[\alpha-\gamma]}-u_{CM})^2+(u_d\sin{[\alpha-\gamma]})^2}
%\label{eq:Urel_dropabs}
%\end{equation}
%\begin{equation}
%U_j^*=\sqrt{(u_j\cos{\gamma}-u_{CM})^2+(u_j\sin{\gamma})^2}
%\label{eq:Urel_jetabs}
%\end{equation}
%Equation (\ref{eq:Urel_abs}) together with the equations (\ref{eq:momentum_xstar}), (\ref{eq:tan_gamma}), (\ref{eq:Urel_dropabs}) and (\ref{eq:Urel_jetabs}) finally leads to the transformation coefficients between the drop and jet relative velocity $U$  and the relative velocities $U_d^*$ and $U_j^*$ of the drop and the jet, respectively, with respect to the centre of mass
%\begin{equation}
%{U_d^*}=\biggl(\frac{m_j}{m_j+m_d}\biggr)\,U
%\label{eq:transf_coeff_Ud}
%\end{equation}
%\begin{equation}
%{U_j^*}=\biggl(1-\frac{m_j}{m_j+m_d}\biggr)\,U
%\label{eq:transf_coeff_Uj}
%\end{equation}
\begin{figure}
\centering
  \includegraphics[width=11cm]{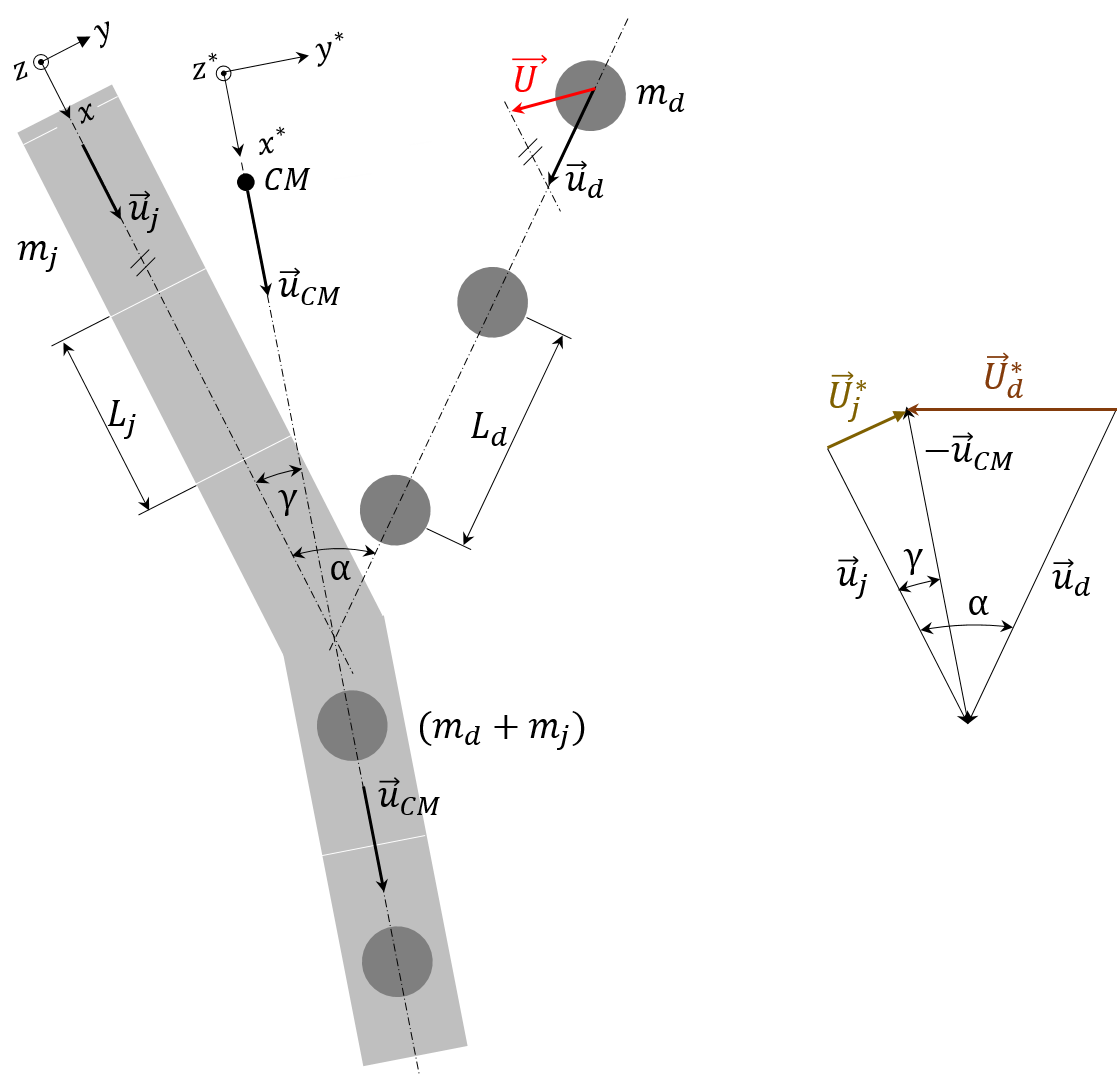}
  \caption{Kinetic parameters of the collision with the $x^*$,$y^*$,$z^*$ reference system placed in the center of mass.}
\label{fig:change_refsystem}
\end{figure}
\FloatBarrier

Thus, the main result of this transformation is that the drop velocity, responsible for the jet deformation, can now be defined relative to the center of mass. Using the relative velocity (in norms):
$U_d^*= m_j/(m_d+m_j)\,U $  with $U$  the relative velocity between the drops and jet (independent from the chosen referential) and $m_j= \rho_j \pi  D_j^2 L_j /4$, $m_d= \rho_d \pi D_d^3/6$ the masses of the jet portion and drop. \\
Having chosen liquids of similar viscosity, we leave for now the viscous losses aside. By doing so, the main forces opposing the drop and therefore the jet deformations are due to surface tensions, dominated at first order by $\sigma_d$. Thus we propose the following  new Weber number: 

\begin{equation}
We_{mom}=\frac{\rho_d D_d {U_d^*}^2}{\sigma_d}=\frac{\rho_d D_d {U}^2}{\sigma_d}\,\biggl(\frac{m_j}{m_d+m_j}\biggr)^2
\label{eq:Weber_d+j}
\end{equation}

\begin{figure}
\centering
  \includegraphics[width=0.5\textwidth]{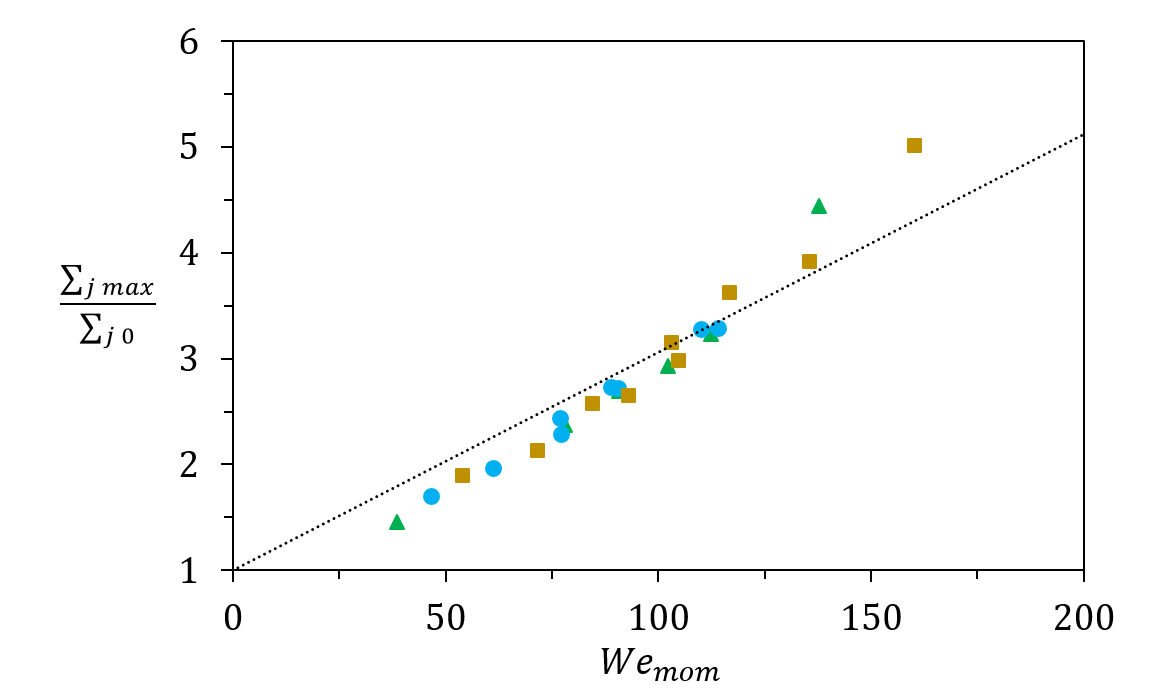}
  \caption{$\Sigma_{j\,max}/ \Sigma_{j{\,}0}$ as a function of $We_{mom}$. Drops of G5 with a jet of SO 5 (green triangles), n-hexadecane (blue circles) and  perfluorodecalin (orange squares). The dotted line  represents eq. (\ref{sigmamax_jet_Wed+j}).}
\label{fig:sigma_jmax_mom}
\end{figure}

We now plot  the normalized jet area at its maximal extension $\Sigma_{j, max}/ \Sigma_{j{,}0}$  as a function of the newly introduced Weber number $We_{mom}$, see figure \ref{fig:sigma_jmax_mom}. First of all, all points collapse on the same curve, while different jet liquids were used with very different densities ranging from about $770$ kg$\cdot$m$^{-3}$ to about $1900$ kg$\cdot$m$^{-3}$. Secondly, the variation is quite well represented by a linear function (dotted line) as 

\begin{equation}
\frac{\Sigma_{j\,max}}{\Sigma_{j{\,}0}}= \alpha_{\Sigma_j} {We_{mom}} + \beta_{\Sigma_j}
\label{sigmamax_jet_Wed+j}
\end{equation}
where $\alpha_{\Sigma_j}=0.0206$ and $\beta_{\Sigma_j}=1$. Deviations appear  for small $We_{mom}$, showing , for $We_d$ approaching zero, the data do not tend as expected towards 1, but toward a smaller value. Deviations are also seen for large $We_{mom}$ ($>120$), corresponding to $We_d>150$, i.e. beyond the studied transition. These deviations could arise from neglected effects which are not negligible any more - especially viscous losses. 
%\FloatBarrier 
%\FloatBarrier 
\section{Temporal evolution of jet or envelope extension}\label{appC}
\begin{figure}
\centering
  \includegraphics[width=0.8\textwidth]{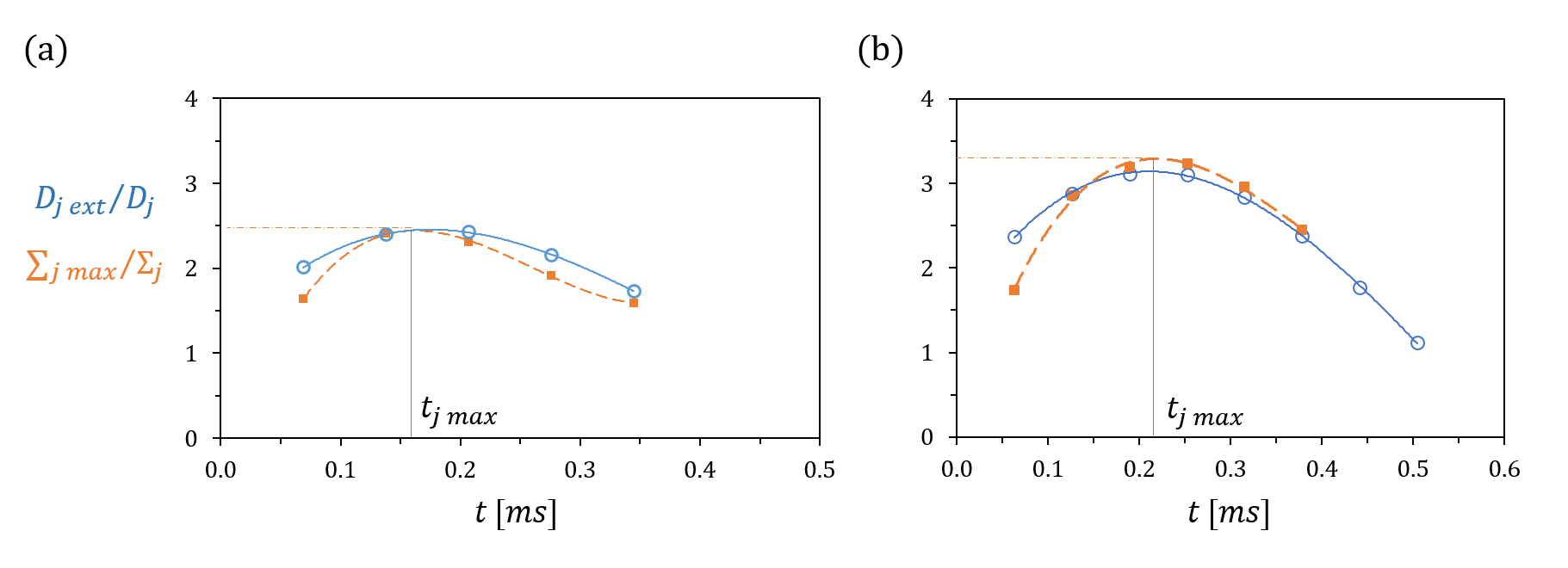}
  \caption{Temporal evolution of  the jet surface $\Sigma_{j\,(t)}/ \Sigma_{j\,0}$ (orange full squares) and of the jet extension $D_{j\,ext(t)}/ D_{j\,0}$ (blue empty circles). The time at which the jet reaches its maximum extension, $t_{j\,max}$, can be equally well determined using the local maximum of $\Sigma_{j\,(t)}/ \Sigma_{j\,0}$ or of $D_{j\,ext(t)}/ D_{j\,0}$ (gray continuous vertical line). $\Sigma_{j\,max}/ \Sigma_{j\,0}$ is represented by the horinzontal orange dashed-pointed lines. The collisions correspond to drops of G5 and a jet of n-hexadecane  with (a) $We_d=120$,  $L_j/D_j=1.45$ and (b) $We_d=176$ and $L_j/D_j=1.9$.}
\label{D_j-max_app}
\end{figure}

In this section, we plot the temporal evaluation of $\Sigma_{j\,(t)}/ \Sigma_{j\,0}$ and $D_{j\,ext(t)}/ D_{j\,0}$, see figure  \ref{D_j-max_app}. Independently from $We_d$ or $L_j/D_j$, we observe that the instant corresponding to the local maximum is very similar for both types of curves. Thus, the fitting procedure applied either to $\Sigma_{j\,(t)}/ \Sigma_{j\,0}$ or to $D_{j\,ext(t)}/ D_{j\,0}$ provides similar results for  $t_{j\,max}$.  Due to possible experimental noise in the evaluation of $\Sigma_{j\,(t)}/ \Sigma_{j\,0}$ (especially for the term $\Sigma_3$ defined in section \ref{sec:image}), we have chosen to use $D_{j\,ext(t)}/ D_{j\,0}$ to obtain $t_{j\, max}$.

\section{Collision parameters and image processing}\label{appD}

This section presents the method used to calculate the drop and jet velocities.

The droplet velocity $\vec{u_d}$ is deduced from the distance separating two consecutive droplets $L_d$ (vectorially $\vec{L_d}$) and the drop frequency $f_d$. The expression reads: $\vec{u_d}=\vec{L_d}\,f_d$, where $f_d$ is set on the signal generator and $\vec{L_d}$ is obtained from the recorded pictures. More precisely, after applying a threshold to separate the drops from the background, the difference between the coordinates of the center of mass of two consecutive drops (given in pixels in the picture referential) provides $\vec{L_d}$ . The norm of the droplet velocity $u_d$ is found to vary between 4 $m.s^{-1}$ and 15 $m.s^{-1}$. 

A similar method is followed for the calculation of the jet velocity $\vec{u_j}$. Thus, before producing the drop-jet collisions, the jet is disturbed by a defined frequency $f_j$ (2000 Hz $<f_j<$ 11000 Hz)  to create a regular stream of droplets of spacial period $L_{d,j}$ (vectorially $\vec{L_{d,j}}$) and diameter $D_{d,j}$. The velocity of these droplets $\vec{u_{d,j}}$ is then obtained as  $\vec{u_{d,j}}=\vec{L_{d,j}}\,f_j$ where $\vec{L_{d,j}}$ is determined from the recorded pictures. The jet supply pressure being kept constant, the volume conservation principle is then applied providing:
$\vec{u_j}=2/3 {D_{d,j}}^3/{D_j}^2f_j$ with $D_j$ the jet diameter measured in the absence of the disturbance.   The norm of the jet velocity $u_j$ is found to vary between 3 $m.s^{-1}$ and 14 $m.s^{-1}$.

\newpage
\bibliographystyle{jfm}

\begin{thebibliography}{71}
\expandafter\ifx\csname natexlab\endcsname\relax\def\natexlab#1{#1}\fi
\def\au#1{#1} \def\ed#1{#1} \def\yr#1{#1}\def\at#1{#1}\def\jt#1{\textit{#1}}
  \def\bt#1{#1}\def\bvol#1{\textbf{#1}} \def\vol#1{#1} \def\pg#1{#1}
  \def\publ#1{#1}\def\arxiv#1{#1}\def\org#1{#1}\def\st#1{\textit{#1}}

\bibitem[sup(2019)]{sup_mat}
 \yr{2019} Supplementary material.

\bibitem[Bazhlekov \& Shopov(1997)]{bazhlekov_shopov_1997}
{\sc \au{Bazhlekov, I.~B.} \& \au{Shopov, P.~J.}} \yr{1997}  \at{Numerical
  simulation of dynamic contact-line problems}.  \jt{J. Fluid Mech.}
  \bvol{352},  \pg{113--133}.

\bibitem[Bird {\em et~al.\/}(2013)Bird, Dhiman, Kwon \& Varanasi]{Bird_2013}
{\sc \au{Bird, J.~C.}, \au{Dhiman, R.}, \au{Kwon, H.-M.} \& \au{Varanasi,
  K.~K.}} \yr{2013}  \at{Reducing the contact time of a bouncing drop}.
  \jt{Nature}  \bvol{503},  \pg{385}.

\bibitem[Blanchette {\em et~al.\/}(2009)Blanchette, Messio \&
  Bush]{Blanchette_2009}
{\sc \au{Blanchette, François}, \au{Messio, Laura} \& \au{Bush, John}}
  \yr{2009}  \at{The influence of surface tension gradients on drop
  coalescence}.  \jt{Physics of Fluids}  \bvol{21}~(072107).

\bibitem[Brandenberger \& Widmer(1998)]{BRANDENBERGER199873}
{\sc \au{Brandenberger, H.} \& \au{Widmer, F.}} \yr{1998}  \at{A new
  multinozzle encapsulation/immobilisation system to produce uniform beads of
  alginate}.  \jt{J. Biotechnol.}  \bvol{63}~(1),  \pg{73 -- 80}.

\bibitem[Brenn {\em et~al.\/}(1996)Brenn, Durst \&
  Tropea]{Brenn-Durst-Tropea_1996}
{\sc \au{Brenn, G.}, \au{Durst, F.} \& \au{Tropea, C.}} \yr{1996}
  \at{Monodisperse sprays for various purposes - their production and
  characteristics}.  \jt{Part. Part. Syst. Charact.}  \bvol{13},
  \pg{179--185}.

\bibitem[Berberovi\ifmmode~\acute{c}\else \'{c}\fi{} {\em
  et~al.\/}(2009)Berberovi\ifmmode~\acute{c}\else \'{c}\fi{}, van Hinsberg,
  Jakirli\ifmmode~\acute{c}\else \'{c}\fi{}, Roisman \& Tropea]{Berberovic2009}
{\sc \au{Berberovi\ifmmode~\acute{c}\else \'{c}\fi{}, E.}, \au{van Hinsberg,
  N.~P.}, \au{Jakirli\ifmmode~\acute{c}\else \'{c}\fi{}, S.}, \au{Roisman,
  I.~V.} \& \au{Tropea, C.}} \yr{2009}  \at{Drop impact onto a liquid layer of
  finite thickness: Dynamics of the cavity evolution}.  \jt{Phys. Rev. E}
  \bvol{79},  \pg{036306}.

\bibitem[Chen {\em et~al.\/}(2017)Chen, Chen \& Amirfazli]{Chen_2017}
{\sc \au{Chen, N.}, \au{Chen, H.} \& \au{Amirfazli, A.}} \yr{2017}  \at{Drop
  impact onto a thin film: Miscibility effect}.  \jt{Phys. Fluids}
  \bvol{29}~(092106).

\bibitem[Chen {\em et~al.\/}(2006)Chen, Chiu \& Lin]{Chen2006_dropjet}
{\sc \au{Chen, R.-H.}, \au{Chiu, S.-L.} \& \au{Lin, T.-H.}} \yr{2006}
  \at{Collisions of a string of water drops on a water jet of equal diameter}.
  \jt{Exp. Therm. Fluid Sci.}  \bvol{31},  \pg{75--81}.

\bibitem[Christen(2010)]{Christen2010}
{\sc \au{Christen, D.S.}} \yr{2010} {\em Praxiswissen der chemischen
  Verfahrenstechnik\/}, 2nd edn. {\em VDI-Buch Chemische Technik /
  Verfahrenstechnik\/} .  \publ{Heidelberg and Dordrecht and London and
  NewYork: Springer}.

\bibitem[Chu {\em et~al.\/}(2007)Chu, Utada, Shah, Kim \& Weitz]{Chu_2007}
{\sc \au{Chu, L.-Y.}, \au{Utada, A.S.}, \au{Shah, R.K.}, \au{Kim, J.-W.} \&
  \au{Weitz, D.A.}} \yr{2007}  \at{Controllable monodisperse multiple
  emulsions}.  \jt{Angew. Chem.}  \bvol{46}~(47),  \pg{8970--8974}.

\bibitem[Cossali {\em et~al.\/}(1997)Cossali, Coghe \& Marengo]{Cossali1997}
{\sc \au{Cossali, G.~E.}, \au{Coghe, A.} \& \au{Marengo, M.}} \yr{1997}
  \at{The impact of a single drop on a wetted solid surface}.  \jt{Exp. Fluids}
   \bvol{22}~(6),  \pg{463--472}.

\bibitem[Dai \& Schmidt(2005)]{Dai2005}
{\sc \au{Dai, M.} \& \au{Schmidt, D.~P.}} \yr{2005}  \at{Numerical simulation
  of head-on droplet collision: Effect of viscosity on maximum deformation}.
  \jt{Phys. Fluids}  \bvol{17},  \pg{041701}.

\bibitem[{De Gennes} {\em et~al.\/}(2004){De Gennes}, Brochard-Wyart \&
  Quere]{Capillarity_2004}
{\sc \au{{De Gennes}, P.~G.}, \au{Brochard-Wyart, F.} \& \au{Quere, D.}}
  \yr{2004} {\em Capillarity and Wetting Phenomena: Drops, Bubbles, Pearls,
  Waves\/}.  \publ{New York: Springer Science}.

\bibitem[Deng {\em et~al.\/}(2013)Deng, Wang, Ju, Xie, Weitz \& Chu]{Deng2013}
{\sc \au{Deng, N.-N.}, \au{Wang, W.}, \au{Ju, X.-J.}, \au{Xie, R.}, \au{Weitz,
  D.~A.} \& \au{Chu, L.-Y.}} \yr{2013}  \at{Wetting-induced formation of
  controllable monodisperse multiple emulsions in microfluidics}.  \jt{Lab
  Chip}  \bvol{13},  \pg{4047--4052}.

\bibitem[Deng {\em et~al.\/}(2014)Deng, Wang, Ju, Xie, Weitz \&
  Chu]{reply_comment_deng_2013}
{\sc \au{Deng, N.-N.}, \au{Wang, W.}, \au{Ju, X.-J.}, \au{Xie, R.}, \au{Weitz,
  D.~A.} \& \au{Chu, L.-Y.}} \yr{2014}  \at{Reply to the ‘comment on
  “wetting-induced formation of controllable monodisperse multiple emulsions
  in microfluidics”’}.  \jt{Lab Chip}  \bvol{14}~(1479).

\bibitem[Einstein(1956)]{Einstein1956}
{\sc \au{Einstein, A.}} \yr{1956} {\em Investigations on the theory of the
  Brownian movement\/}.  \publ{New York, N.Y.: Dover Publications, Inc.}

\bibitem[Gao {\em et~al.\/}(2005)Gao, Chen, Pu \& Lin]{Gao2005}
{\sc \au{Gao, T.-C.}, \au{Chen, R.-H.}, \au{Pu, J.-Y.} \& \au{Lin, T.-H.}}
  \yr{2005}  \at{Collision between an ethanol drop and a water drop}.  \jt{Exp.
  Fluids}  \bvol{38}~(6),  \pg{731--738}.

\bibitem[Georgiev {\em et~al.\/}(2018)Georgiev, Danov, Kralchevsky, Gurkov,
  Krusteva, Arnaudov, Stoyanov \& Pelan]{Georgie2018}
{\sc \au{Georgiev, M.~T.}, \au{Danov, K.~D.}, \au{Kralchevsky, P.~A.},
  \au{Gurkov, T.~D.}, \au{Krusteva, D.~P.}, \au{Arnaudov, L.~N.}, \au{Stoyanov,
  S.~D.} \& \au{Pelan, E.~G.}} \yr{2018}  \at{Rheology of particle/water/oil
  three-phase dispersions: Electrostatic vs. capillary bridge forces}.  \jt{J.
  Colloid Interface Sci.}  \bvol{513},  \pg{515 -- 526}.

\bibitem[Geppert {\em et~al.\/}(2017)Geppert, Terzis, Lamanna, Marengo \&
  Weigand]{Geppert_2017}
{\sc \au{Geppert, Anne}, \au{Terzis, Aris}, \au{Lamanna, Giovanni},
  \au{Marengo, Marco} \& \au{Weigand, Bernhard}} \yr{2017}  \at{A benchmark
  study for the crown-type splashing dynamics of one- and two-component droplet
  wall–film interactions}.  \jt{Experiments in Fluids}  \bvol{58},
  \pg{1--27}.

\bibitem[Gotaas {\em et~al.\/}(2007)Gotaas, Havelka, Jakobsen, Svendsen, Hase,
  Roth \& Weigand]{Weigand2007}
{\sc \au{Gotaas, C.}, \au{Havelka, P.}, \au{Jakobsen, H.~A.}, \au{Svendsen,
  H.~F.}, \au{Hase, M.}, \au{Roth, N.} \& \au{Weigand, B.}} \yr{2007}
  \at{Effect of viscosity on droplet-droplet collision outcome: Experimental
  study and numerical comparison}.  \jt{Phys. Fluids}  \bvol{19}~(10),
  \pg{102\--106}.

\bibitem[Guzowski \& Garstecki(2014)]{comment_deng_2013}
{\sc \au{Guzowski, J.} \& \au{Garstecki, P.}} \yr{2014}  \at{Comment on
  “wetting-induced formation of controllable monodisperse multiple emulsions
  in microfluidics”}.  \jt{Lab Chip}  \bvol{14}~(1477).

\bibitem[Haeberle {\em et~al.\/}(2008)Haeberle, Naegele, Burger, von Stetten,
  Zengerle \& Ducrée]{Haeberle2008}
{\sc \au{Haeberle, S.}, \au{Naegele, L.}, \au{Burger, R.}, \au{von Stetten,
  F.}, \au{Zengerle, R.} \& \au{Ducrée, J.}} \yr{2008}  \at{Alginate bead
  fabrication and encapsulation of living cells under centrifugally induced
  artificial gravity conditions}.  \jt{J. Microencapsul.}  \bvol{25}~(4),
  \pg{267--274}.

\bibitem[Hinterbichler {\em et~al.\/}(2015)Hinterbichler, Planchette \&
  Brenn]{Hinterbichler-Planchette-Brenn_2015}
{\sc \au{Hinterbichler, H.}, \au{Planchette, C.} \& \au{Brenn, G.}} \yr{2015}
  \at{Ternary drop collisions}.  \jt{Exp. Fluids}  \bvol{56}~(190).

\bibitem[Hoath {\em et~al.\/}(2013)Hoath, Jung \& Hutchings]{hoath_2013}
{\sc \au{Hoath, S.~D.}, \au{Jung, S.} \& \au{Hutchings, I.~M.}} \yr{2013}
  \at{A simple criterion for filament break-up in drop-on-demand inkjet
  printing}.  \jt{Phys. Fluids}  \bvol{25}~(2),  \pg{021701}.

\bibitem[Hoepffner \& Paré(2013)]{hoepffner2013}
{\sc \au{Hoepffner, J.} \& \au{Paré, G.}} \yr{2013}  \at{Recoil of a liquid
  filament: escape from pinch-off through creation of a vortex ring}.  \jt{J.
  Fluid Mech.}  \bvol{734},  \pg{183–--197}.

\bibitem[Jiang {\em et~al.\/}(1992)Jiang, Umemura \&
  Law]{Jiang-Umemura-Law_1992}
{\sc \au{Jiang, Y.~J.}, \au{Umemura, A.} \& \au{Law, C.~K.}} \yr{1992}  \at{An
  experimental investigation in the collision behaviour of hydrocarbon
  droplets}.  \jt{J. Fluid Mech.}  \bvol{234},  \pg{171--190}.

\bibitem[Josserand \& Thoroddsen(2016)]{josserand2016}
{\sc \au{Josserand, C.} \& \au{Thoroddsen, S.~T.}} \yr{2016}  \at{Drop impact
  on a solid surface}.  \jt{{Annu. Rev. Fluid Mech.}}  \bvol{48},
  \pg{365--391}.

\bibitem[Josserand \& Zaleski(2003)]{Josserand2003}
{\sc \au{Josserand, C.} \& \au{Zaleski, S.}} \yr{2003}  \at{Droplet splashing
  on a thin liquid film}.  \jt{Phys. Fluids}  \bvol{15}~(6),  \pg{1650--1657}.

\bibitem[Kadota \& Yamasaki(2002)]{KADOTA_2002}
{\sc \au{Kadota, T.} \& \au{Yamasaki, H.}} \yr{2002}  \at{Recent advances in
  the combustion of water fuel emulsion}.  \jt{Progress in Energy and
  Combustion Science}  \bvol{28}~(5),  \pg{385 -- 404}.

\bibitem[Kamperman {\em et~al.\/}(2018)Kamperman, Trikalitis, Karperien, Visser
  \& Leijten]{Kamperman2018}
{\sc \au{Kamperman, T.}, \au{Trikalitis, V.~D.}, \au{Karperien, M.},
  \au{Visser, C.~W.} \& \au{Leijten, J.}} \yr{2018}  \at{Ultrahigh-throughput
  production of monodisperse and multifunctional janus microparticles using
  in-air microfluidics}.  \jt{ACS Appl. Mater. Interfaces}  \bvol{10}~(28),
  \pg{23433--23438}.

\bibitem[Kavehpour(2015)]{Kavehpour_2014}
{\sc \au{Kavehpour, H.~P.}} \yr{2015}  \at{Coalescence of drops}.  \jt{Annu.
  Rev. Fluid Mech.}  \bvol{47}~(1),  \pg{245--268}.

\bibitem[Khademhosseini {\em et~al.\/}(2006)Khademhosseini, Langer, Borenstein
  \& Vacanti]{Khademhosseini2480}
{\sc \au{Khademhosseini, A.}, \au{Langer, R.}, \au{Borenstein, J.} \&
  \au{Vacanti, J.~P.}} \yr{2006}  \at{Microscale technologies for tissue
  engineering and biology}.  \jt{Proc. Natl. Acad. Sci. U.S.A.}
  \bvol{103}~(8),  \pg{2480--2487}.

\bibitem[Kittel {\em et~al.\/}(2018)Kittel, Roisman \&
  Tropea]{Kittel_Roisman2018}
{\sc \au{Kittel, H.~M.}, \au{Roisman, I.~V.} \& \au{Tropea, C.}} \yr{2018}
  \at{Splash of a drop impacting onto a solid substrate wetted by a thin film
  of another liquid}.  \jt{Phys. Rev. Fluids}  \bvol{3},  \pg{073601}.

\bibitem[Kralchevsky(2019)]{kralchevsky_2019}
{\sc \au{Kralchevsky, P.A.}} \yr{2019}  \at{private discussion}  \bvol{Graz
  (AT) and Sofia (BG)}.

\bibitem[Lhuissier {\em et~al.\/}(2013)Lhuissier, Sun, Prosperetti \&
  Lohse]{Lhuissier2013}
{\sc \au{Lhuissier, H.}, \au{Sun, C.}, \au{Prosperetti, A.} \& \au{Lohse, D.}}
  \yr{2013}  \at{Drop fragmentation at impact onto a bath of an immiscible
  liquid}.  \jt{Phys. Rev. Lett}  \bvol{110},  \pg{264503}.

\bibitem[Liu {\em et~al.\/}(2018)Liu, Zhang, Gao, Lu \&
  Ding]{liu_zhang_gao_lu_ding_2018}
{\sc \au{Liu, H.-R.}, \au{Zhang, C.-Y.}, \au{Gao, P.}, \au{Lu, X.-Y.} \&
  \au{Ding, H.}} \yr{2018}  \at{On the maximal spreading of impacting compound
  drops}.  \jt{J. Fluid Mech.}  \bvol{854},  \pg{R6}.

\bibitem[Lunkad {\em et~al.\/}(2007)Lunkad, Buwa \& Nigam]{LUNKAD20077214}
{\sc \au{Lunkad, S.~F.}, \au{Buwa, V.~V.} \& \au{Nigam, K.D.P.}} \yr{2007}
  \at{Numerical simulations of drop impact and spreading on horizontal and
  inclined surfaces}.  \jt{Chem. Eng. Sci.}  \bvol{62}~(24),  \pg{7214 --
  7224}, 8th International Conference on Gas-Liquid and Gas-Liquid-Solid
  Reactor Engineering.

\bibitem[Marangon {\em et~al.\/}(2019)Marangon, Hsiao, Brenn \&
  Planchette]{ilass_2019}
{\sc \au{Marangon, F.}, \au{Hsiao, W.K.}, \au{Brenn, G.} \& \au{Planchette,
  C.}} \yr{2019}  \at{Satellite drop formation during piezo-based inkjet
  printing}.  \jt{Proceedings of the 29th Conference on Liquid Atomization and
  Spray Systems, September 2019, Paris, France} .

\bibitem[Martin {\em et~al.\/}(2008)Martin, Hoath \& Hutchings]{Martin_2008}
{\sc \au{Martin, G.~D.}, \au{Hoath, S.~D.} \& \au{Hutchings, I.~M.}} \yr{2008}
  \at{Inkjet printing - the physics of manipulating liquid jets and drops}.
  \jt{J. Phys. Conf. Ser.}  \bvol{105},  \pg{012001}.

\bibitem[Mazloomi {\em et~al.\/}(2016)Mazloomi, Chikatamarla \&
  Karlin]{Mazloomi2016}
{\sc \au{Mazloomi, A.}, \au{Chikatamarla, S.} \& \au{Karlin, I.}} \yr{2016}
  \at{Simulation of binary droplet collisions with the entropic lattice
  boltzmann method}.  \jt{Phys. Fluids}  \bvol{28},  \pg{022106}.

\bibitem[Okumura {\em et~al.\/}(2003)Okumura, Chevy, Richard, Qu\'{e}r\'{e} \&
  Clanet]{Okumura2003}
{\sc \au{Okumura, K.}, \au{Chevy, F.}, \au{Richard, D.}, \au{Qu\'{e}r\'{e}, D.}
  \& \au{Clanet, C.}} \yr{2003}  \at{Water spring: A model for bouncing drops}.
   \jt{Europhys. Lett.}  \bvol{62}~(2),  \pg{237--243}.

\bibitem[Planchette {\em et~al.\/}(2018{\natexlab{{\em a\/}}})Planchette,
  Hinterbichler \& Brenn]{Planchette2018}
{\sc \au{Planchette, C.}, \au{Hinterbichler, H.} \& \au{Brenn, G.}}
  \yr{2018{\natexlab{{\em a\/}}}}  \at{Drop stream - immiscible jet collisions:
  Regimes and fragmentation mechanisms}.  \jt{Proceedings of the 28th
  Conference on Liquid Atomization and Spray Systems, 6-8 September 2017,
  Valencia, Spain}  \pg{p.~7}.

\bibitem[Planchette {\em et~al.\/}(2017)Planchette, Hinterbichler, Liu, Bothe
  \& Brenn]{PlanchetteBrenn2017}
{\sc \au{Planchette, C.}, \au{Hinterbichler, H.}, \au{Liu, M.}, \au{Bothe, D.}
  \& \au{Brenn, G.}} \yr{2017}  \at{Colliding drops as coalescing and
  fragmenting liquid springs}.  \jt{J. Fluid Mech.}  \bvol{814},
  \pg{277--300}.

\bibitem[Planchette {\em et~al.\/}(2010)Planchette, Lorenceau \&
  Brenn]{PLANCHETTE201089}
{\sc \au{Planchette, C.}, \au{Lorenceau, E.} \& \au{Brenn, G.}} \yr{2010}
  \at{Liquid encapsulation by binary collisions of immiscible liquid drops}.
  \jt{Colloids Surf. A}  \bvol{365}~(1),  \pg{89 -- 94}.

\bibitem[Planchette {\em et~al.\/}(2012)Planchette, Lorenceau \&
  Brenn]{Planchette2012}
{\sc \au{Planchette, C.}, \au{Lorenceau, E.} \& \au{Brenn, G.}} \yr{2012}
  \at{The onset of fragmentation in binary liquid drop collisions}.  \jt{J.
  Fluid Mech.}  \bvol{702},  \pg{5--25}.

\bibitem[Planchette {\em et~al.\/}(2018{\natexlab{{\em b\/}}})Planchette,
  Petit, Hinterbichler \& Brenn]{PlanchetteBrenn2018}
{\sc \au{Planchette, C.}, \au{Petit, S.}, \au{Hinterbichler, H.} \& \au{Brenn,
  G.}} \yr{2018{\natexlab{{\em b\/}}}}  \at{Collisions of drops with an
  immiscible liquid jet}.  \jt{Phys. Rev. Fluids}  \bvol{3},  \pg{093603}.

\bibitem[Rein(1993)]{Rein1993}
{\sc \au{Rein, M.}} \yr{1993}  \at{Phenomena of liquid drop impact on solid and
  liquid surfaces}.  \jt{Fluid Dyn. Res.}  \bvol{12}~(2),  \pg{61--93}.

\bibitem[Richard {\em et~al.\/}(2002)Richard, Clanet \&
  Qu\'{e}r\'{e}]{Richard2002}
{\sc \au{Richard, D.}, \au{Clanet, C.} \& \au{Qu\'{e}r\'{e}, D.}} \yr{2002}
  \at{Contact time of a bouncing drop}.  \jt{Nature}  \bvol{417},  \pg{811}.

\bibitem[Roisman(2004)]{Roisman2004}
{\sc \au{Roisman, I.~V.}} \yr{2004}  \at{Dynamics of inertia dominated binary
  drop collisions}.  \jt{Phys. Fluids}  \bvol{16}~(9),  \pg{3438--3449}.

\bibitem[Ross \& Becher(1992)]{Ross1992}
{\sc \au{Ross, S.} \& \au{Becher, P.}} \yr{1992}  \at{The history of the
  spreading coefficient}.  \jt{J. Colloid Interface Sci.}  \bvol{149}~(2),
  \pg{575--579}.

\bibitem[Santiago-Rosanne {\em et~al.\/}(2001)Santiago-Rosanne, Vignes-Adler \&
  Velarde]{SANTIAGOROSANNE2001375}
{\sc \au{Santiago-Rosanne, Maria}, \au{Vignes-Adler, Michèle} \& \au{Velarde,
  Manuel~G.}} \yr{2001}  \at{On the spreading of partially miscible liquids}.
  \jt{Journal of Colloid and Interface Science}  \bvol{234}~(2),  \pg{375 --
  383}.

\bibitem[Schroll {\em et~al.\/}(2010)Schroll, Josserand, Zaleski \&
  Zhang]{Schroll_Zaleski2010}
{\sc \au{Schroll, R.~D.}, \au{Josserand, C.}, \au{Zaleski, S.} \& \au{Zhang,
  W.~W.}} \yr{2010}  \at{Impact of a viscous liquid drop}.  \jt{Phys. Rev.
  Lett.}  \bvol{104},  \pg{034504}.

\bibitem[Serp {\em et~al.\/}(2000)Serp, Cantana, Heinzen, Von~Stockar \&
  Marison]{Serp2000}
{\sc \au{Serp, D.}, \au{Cantana, E.}, \au{Heinzen, C.}, \au{Von~Stockar, U.} \&
  \au{Marison, I.~W.}} \yr{2000}  \at{Characterization of an encapsulation
  device for the production of monodisperse alginate beads for cell
  immobilization}.  \jt{Biotechnol. Bioeng.}  \bvol{70}~(1),  \pg{41--53}.

\bibitem[Shikhmurzaev(2008)]{shikhmurzaev}
{\sc \au{Shikhmurzaev, Y.~D.}} \yr{2008} {\em Capillary flows with forming
  interfaces\/}.  \publ{Boca Raton, FL, USA: CRC Press, Chapman and Hall}.

\bibitem[Stone {\em et~al.\/}(1986)Stone, Bentley \& Leal]{Stone1986}
{\sc \au{Stone, H.~A.}, \au{Bentley, B.~J.} \& \au{Leal, L.~G.}} \yr{1986}
  \at{An experimental study of transient effects in the breakup of viscous
  drops}.  \jt{J. Fluid Mech.}  \bvol{173},  \pg{131--158}.

\bibitem[Stone \& Leal(1989{\natexlab{{\em a\/}}})]{stone_leal_1989}
{\sc \au{Stone, H.~A.} \& \au{Leal, L.~G.}} \yr{1989{\natexlab{{\em a\/}}}}
  \at{The influence of initial deformation on drop breakup in subcritical
  time-dependent flows at low reynolds numbers}.  \jt{Journal of Fluid
  Mechanics}  \bvol{206},  \pg{223–263}.

\bibitem[Stone \& Leal(1989{\natexlab{{\em b\/}}})]{Stone1989}
{\sc \au{Stone, H.~A.} \& \au{Leal, L.~G.}} \yr{1989{\natexlab{{\em b\/}}}}
  \at{Relaxation and breakup of an initially extended drop in an otherwise
  quiescent fluid}.  \jt{J. Fluid Mech.}  \bvol{198},  \pg{399--427}.

\bibitem[Sui {\em et~al.\/}(2014)Sui, Ding \& Spelt]{Sui2014}
{\sc \au{Sui, Y.}, \au{Ding, H.} \& \au{Spelt, P.D.M.}} \yr{2014}
  \at{Numerical simulations of flows with moving contact lines}.  \jt{Annu.
  Rev. Fluid Mech.}  \bvol{46}~(1),  \pg{97--119}.

\bibitem[Takamura {\em et~al.\/}(2012)Takamura, Fischer \&
  Morrow]{TAKAMURA201250}
{\sc \au{Takamura, K.}, \au{Fischer, H.} \& \au{Morrow, N.~R.}} \yr{2012}
  \at{Physical properties of aqueous glycerol solutions}.  \jt{J. Petrol. Sci.
  Eng.}  \bvol{98-99},  \pg{50--60}.

\bibitem[Teh {\em et~al.\/}(2008)Teh, Lin, Hung \& Lee]{Teh2008}
{\sc \au{Teh, S.-Y.}, \au{Lin, R.}, \au{Hung, L.-H.} \& \au{Lee, A.P.}}
  \yr{2008}  \at{Droplet microfluidics}.  \jt{Lab Chip}  \bvol{8},
  \pg{198--220}.

\bibitem[Tsuru(2019)]{Tsuru_2019}
{\sc \au{Tsuru, Daisuke}} \yr{2019}  \at{Private discussion}  \bvol{Graz}.

\bibitem[Visser {\em et~al.\/}(2018)Visser, Kamperman, Karbaat, Lohse \&
  Karperien]{Visser2018}
{\sc \au{Visser, C.~W.}, \au{Kamperman, T.}, \au{Karbaat, L.~P.}, \au{Lohse,
  D.} \& \au{Karperien, M.}} \yr{2018}  \at{In-air microfluidics enables rapid
  fabrication of emulsions, suspensions, and 3d modular (bio)materials}.
  \jt{Sci. Adv}  \bvol{4}~(1).

\bibitem[Wang \& Chen(2000)]{Wang2000}
{\sc \au{Wang, A.-B.} \& \au{Chen, C.-C.}} \yr{2000}  \at{Splashing impact of a
  single drop onto very thin liquid films}.  \jt{Phys. Fluids}  \bvol{12}~(9),
  \pg{2155--2158}.

\bibitem[Wang {\em et~al.\/}(2004)Wang, Lin, Hung, Huang \& Law]{Wang_2004}
{\sc \au{Wang, C.~H.}, \au{Lin, C.~Z.}, \au{Hung, W.~G.}, \au{Huang, W.~C.} \&
  \au{Law, C.~K.}} \yr{2004}  \at{On the burning characteristics of
  collision-generated water/hexadecane droplets}.  \jt{Combust. Sci. Technol.}
  \bvol{176}~(1),  \pg{71--96}.

\bibitem[Wildeman {\em et~al.\/}(2016)Wildeman, Visser, Sun \&
  Lohse]{Wildeman2016}
{\sc \au{Wildeman, S.}, \au{Visser, C.~W.}, \au{Sun, C.} \& \au{Lohse, D.}}
  \yr{2016}  \at{On the spreading of impacting drops}.  \jt{J. Fluid Mech.}
  \bvol{805},  \pg{636--655}.

\bibitem[Willis \& Orme(2003)]{Willis-Orme_2003}
{\sc \au{Willis, K.~D.} \& \au{Orme, M.}} \yr{2003}  \at{Binary droplet
  collisions in a vacuum environment: an experimental investigation on the role
  of viscosity}.  \jt{Exp. Fluids}  \bvol{34},  \pg{28--41}.

\bibitem[W\"ohrwag {\em et~al.\/}(2018)W\"ohrwag, Semprebon, Mazloomi~Moqaddam,
  Karlin \& Kusumaatmaja]{semprebon_2018}
{\sc \au{W\"ohrwag, M.}, \au{Semprebon, C.}, \au{Mazloomi~Moqaddam, A.},
  \au{Karlin, I.} \& \au{Kusumaatmaja, H.}} \yr{2018}  \at{Ternary free-energy
  entropic lattice {B}oltzmann model with a high density ratio}.  \jt{Phys.
  Rev. Lett.}  \bvol{120},  \pg{234501}.

\bibitem[Yarin(2006)]{Yarin2006}
{\sc \au{Yarin, A.L.}} \yr{2006}  \at{Drop impact dynamics: Splashing,
  spreading, receding, bouncing…}.  \jt{Annu. Rev. Fluid Mech.}
  \bvol{38}~(1),  \pg{159--192}.

\bibitem[Yarin {\em et~al.\/}(2014)Yarin, Pourdeyhimi \& Ramakrishna]{fibers}
{\sc \au{Yarin, A.~L.}, \au{Pourdeyhimi, B.} \& \au{Ramakrishna, S.}} \yr{2014}
  {\em Fundamentals and Applications of Micro- and Nanofibers\/}.
  \publ{Cambridge University Press}.

\bibitem[Yeo {\em et~al.\/}(2004)Yeo, Chen, Basaran \&
  Park]{Yeo_Chen_Basaran_Park_2004}
{\sc \au{Yeo, Y.}, \au{Chen, A.~U.}, \au{Basaran, O.~A.} \& \au{Park, K.}}
  \yr{2004}  \at{Solvent exchange method: a novel microencapsulation technique
  using dual microdispensers}.  \jt{Pharm. Res.}  \bvol{21},  \pg{1419--1427}.

\end{thebibliography}

\end{document}